\documentclass[aps,11pt,prd,longbibliography]{revtex4-1}
\usepackage[most]{tcolorbox}
\usepackage{slashed}

\usepackage{tikz}
\usepackage{feynmp-auto}
\usepackage{amsmath}
\usepackage[utf8]{inputenc}
\usepackage[T1]{fontenc} 
\usepackage{array}
\usepackage{bbm}
\usepackage{soul}	
\usepackage{xcolor}

\usepackage{epsfig}
\usepackage{color}
\usepackage{slashed}
\usepackage{comment}
\usepackage{epstopdf}
\usepackage{xspace}
\usepackage{mathrsfs}
\usepackage[euler]{textgreek}
\usepackage{amsmath}
\usepackage{amsthm}
\usepackage{amsfonts}
\usepackage{amssymb}
\usepackage{graphicx}
\graphicspath{ {Figures/} }
\usepackage[font={small}]{caption}
\usepackage{subcaption}
\usepackage{float}
\usepackage{multirow}
\usepackage[hang, flushmargin,bottom]{footmisc} 
\usepackage{placeins}
\usepackage{enumitem}
\usepackage{slashed}
\usepackage{bbm}
\usepackage{appendix}
\usepackage{tabularx}
\usepackage{siunitx}

\usepackage{titlesec}
\titleformat{\chapter}[display]
  {\normalfont\LARGE\bfseries}
  {\chaptertitlename\ \thechapter}{5pt}{\LARGE}
  \titlespacing*{\chapter}{0pt}{-20pt}{35pt}
\usepackage{bigstrut}
\usepackage{pst-node}
\usepackage{pstricks}
\usepackage{physics}
\usepackage{mathtools}
\usepackage{setspace}
\usepackage{fancyhdr}
\usepackage[makeroom]{cancel}
\usepackage{feyn}
\newcommand{\be}{\begin{equation}}
\newcommand{\ee}{\end{equation}}
\newcommand{\bes}{\begin{equation*}}
\newcommand{\ees}{\end{equation*}}

\usepackage{hyperref}
\hypersetup{%
  colorlinks = true,
  linkcolor  = blue,
  citecolor  = blue,
  urlcolor   = blue,
}
\usepackage{soul}
\usepackage{todonotes}
\usepackage{xpatch}
\makeatletter
\xpretocmd{\todo}{\@bsphack}{}{}
\xapptocmd{\todo}{\@esphack}{}{}
\makeatother

\newcommand{\beq}{\begin{equation}}
\newcommand{\eeq}{\end{equation}}

\newcommand{\gB}{\ensuremath{g_\mathrm{B}}\xspace}

\newcommand{\ZB}{\ensuremath{Z_\mathrm{B}}\xspace}
\newcommand{\HB}{\ensuremath{h_\mathrm{B}}\xspace}

\newcommand{\tB}{\ensuremath{\theta_\mathrm{B}}\xspace}

\definecolor{green}{HTML}{008000}
\definecolor{goldenrod}{HTML}{DAA520}
\definecolor{magenta}{HTML}{FF00FF}
\definecolor{silver}{HTML}{C0C0C0}
\definecolor{indigo}{HTML}{4B0082}
\definecolor{skyblue}{HTML}{87CEEB}
\definecolor{darkgoldenrod}{HTML}{B8860B}
\definecolor{orange}{HTML}{FFA500}
\definecolor{yellow}{HTML}{FFFF00}
\definecolor{saddlebrown}{HTML}{8B4513}
\definecolor{blue}{HTML}{0000FF}
\definecolor{turquoise}{HTML}{40E0D0}
\definecolor{yellow}{HTML}{FFFF00}
\definecolor{white}{HTML}{FFFFFF}
\definecolor{whitesmoke}{HTML}{F5F5F5}

\newcommand{\myComment}[1]{}

\usepackage{xparse}

\usetikzlibrary{decorations.pathmorphing}	
\usetikzlibrary{decorations.markings}
\tikzset{
    vector/.style={decorate, decoration={snake}, draw},
	provector/.style={decorate, decoration={snake,amplitude=2.5pt}, draw},
	antivector/.style={decorate, decoration={snake,amplitude=-2.5pt}, draw},
    fermion/.style={draw=black, postaction={decorate},
        decoration={markings,mark=at position .55 with {\arrow[draw=black]{>}}}},
    fermionr/.style={draw=black, postaction={decorate},
    decoration={markings,mark=at position .55 with {\arrow[draw=black]{<}}}},
    fermioncyan/.style={draw=black, postaction={decorate},
        decoration={markings,mark=at position .55 with {\arrow[draw=cyan]{<}}}},
    fermiondif/.style={draw=black, postaction={decorate},
        decoration={markings,mark=at position .7 with {\arrow[draw=black]{>}}}},
            fermiondif2/.style={draw=black, postaction={decorate},
        decoration={markings,mark=at position .7 with {\arrow[draw=black]{<}}}},
    fermionend/.style={draw=black, postaction={decorate},
        decoration={markings,mark=at position 1 with {\arrow[draw=black]{>}}}},
    fermionuchannel2/.style={draw=black, postaction={decorate},
        decoration={markings,mark=at position .4 with {\arrow[draw=black]{>}}}},
    scalardif/.style={dashed,draw=black, postaction={decorate},
        decoration={markings,mark=at position .7 with {\arrow[draw=black]{>}}}},
    scalarend/.style={dashed,draw=black, postaction={decorate},
        decoration={markings,mark=at position 1 with {\arrow[draw=black]{>}}}},
    fermionbar/.style={draw=black, postaction={decorate},
        decoration={markings,mark=at position .55 with {\arrow[draw=black]{<}}}},
    fermionnoarrow/.style={draw=black},
    gluon/.style={decorate, draw=black,
        decoration={coil,amplitude=4pt, segment length=5pt}},
    scalar/.style={dashed,draw=black, postaction={decorate},
        decoration={markings,mark=at position .55 with {\arrow[draw=black]{>}}}},
    scalarcyan/.style={dashed,draw=black, postaction={decorate},
        decoration={markings,mark=at position .55 with {\arrow[draw=cyan]{>}}}},
    scalaruchannel1/.style={dashed,draw=black, postaction={decorate},
        decoration={markings,mark=at position .7 with {\arrow[draw=black]{>}}}},
                  scalaruchannel2/.style={dashed,draw=black, postaction={decorate},
        decoration={markings,mark=at position .4 with {\arrow[draw=black]{>}}}},
    scalarbar/.style={dashed,draw=black, postaction={decorate},
        decoration={markings,mark=at position .55 with {\arrow[draw=black]{<}}}},
    scalarnoarrow/.style={dashed,draw=black},
    electron/.style={draw=black, postaction={decorate},
        decoration={markings,mark=at position .55 with {\arrow[draw=black]{>}}}},
	bigvector/.style={decorate, decoration={snake,amplitude=4pt}, draw},
}

\NewDocumentCommand\semiloop{O{black}mmmO{}O{above}}
{%
\draw[#1] let \p1 = ($(#3)-(#2)$) in (#3) arc (#4:({#4+180}):({0.5*veclen(\x1,\y1)})node[midway, #6] {#5};)
}

\tikzstyle{block} = [draw, rectangle, 
    minimum height=3em, minimum width=6em]

\usepackage{enumitem}
\usepackage{tikz}
\usetikzlibrary{fit}
\tikzset{%
  highlight/.style={rectangle,rounded corners,color=granate,draw,text opacity =1,
    fill opacity=0.5,thick,inner sep=0pt}
}

\usepackage{xparse}
\NewDocumentCommand\loopv{O{black}mmmO{}O{above}}
{%
\draw[#1] let \p1 = ($(#3)-(#2)$) in (#3) arc (#4:({#4+360}):({0.5*veclen(\x1,\y1)})node[midway, #6] {#5};)
}

\usepackage{placeins}

\tikzset{
    cross/.pic = {
    \draw[rotate = 45] (-#1,0) -- (#1,0);
    \draw[rotate = 45] (0,-#1) -- (0, #1);
    }
}

\tikzset{
    square/.style={%
        draw=none,
        circle,
        append after command={%
            \pgfextra \draw[#1] (\tikzlastnode.north-|\tikzlastnode.west) rectangle 
                (\tikzlastnode.south-|\tikzlastnode.east);\endpgfextra}
    },
    square/.default=black
}

\tikzstyle{block} = [draw, rectangle, 
    minimum height=3em, minimum width=6em]
\begin{document}
\title{\Large{Gamma Lines and Dark Matter from Anomaly Cancellation}}
\author{Hridoy Debnath, Pavel Fileviez P\'erez, Kevin Gonz\'alez-Quesada}
\affiliation{
Physics Department and Center for Education and Research in Cosmology and Astrophysics (CERCA), Case Western Reserve University, Cleveland, OH 44106, USA}
\email{hxd253@case.edu, pxf112@case.edu, kag155@case.edu}
\date{\today}
\begin{abstract}
We discuss a simple theory for physics beyond the Standard Model where a Majorana dark matter is predicted from anomaly cancellation. We discuss in detail the minimal theory where the baryon number is a local symmetry spontaneously broken at the low scale. The correlation between the cosmological constraints on the dark matter relic density, the direct detection and collider bounds is investigated. We discuss in great detail the gamma lines from dark matter annihilation showing the possibility to test these predictions in the near future at gamma-ray telescopes such as CTA. We investigate all processes contributing to the total photon flux from dark matter annihilation and point out the unique features that can be used to test this theory for dark matter.  
\end{abstract}

\maketitle
\section{INTRODUCTION}
%
The dark matter problem is one of the most pressing issues in modern cosmology. Today, we know about dark matter candidates that could be discovered in the near future~\cite{Freese:2017idy,Drees:2018hzm}.
Unfortunately, the standard model of particle physics does not provide a dark matter candidate and one needs to look for new theories beyond the standard model where one could predict the existence of a dark matter candidate. The particle physics community believed in the ideas related to Supersymmetry and studied carefully the lightest neutralino as a cold dark matter candidate~\cite{Jungman:1995df}. The QCD axion has been also a very popular dark matter candidate, see Ref.~\cite{DiLuzio:2020wdo} for a review. It is fair to say that the weakly interacting massive particles and the QCD axion have been the popular dark matter candidates in the particle physics community, because they are predicted in extensions of the standard model where one address some of the theoretical issues in the standard model.

In this article, we investigate a class of theories for new physics where a cold dark matter candidate is predicted from gauge anomaly cancellation. The standard model of particle physics is an anomaly-free gauge theory and the fermionic degrees of freedom are exactly the needed for anomaly cancellation. Anomaly cancellation can be seen also as a theoretical principle that can help us to look for new gauge theories beyond the standard model. One can imagine many new gauge symmetries but the standard model has two types of accidental
global symmetries at the classical level~\footnote{These symmetries are broken at the quantum level by $SU(2)_L$ instantons~\cite{tHooft:1976rip}.}, the total lepton number and baryon number, that could
be promoted to local gauge symmetries~\cite{FileviezPerez:2010gw,FileviezPerez:2011pt,Duerr:2013dza,FileviezPerez:2014lnj,FileviezPerez:2024fzc}. These theories predict a dark matter candidate from anomaly cancellation and make several interesting predictions for collider and dark matter experiments. See Ref.~\cite{FileviezPerez:2024fzc} for a new proposal where the gauge anomalies associated to lepton or baryon numbers are cancelled with the minimal number of extra fermionic representations and predict a cold dark matter candidate. These theories are interesting because they are motivated for other reasons and as a bonus one predicts the possible existence of a fermionic dark matter candidate.

Theories where a dark matter candidate is predicted from the cancellation of leptonic or baryonic anomalies have been studied in Refs.~\cite{Duerr:2013lka,Duerr:2014wra,Duerr:2015vna,FileviezPerez:2018jmr,FileviezPerez:2019jju,FileviezPerez:2019cyn,Debnath:2023akj,Debnath:2024vpf,Butterworth:2024eyr}. See also the study in Ref.~\cite{Ellis:2017tkh}. In this article, we investigate the properties of the fermionic dark matter candidates in the minimal theory for local baryon number~\cite{FileviezPerez:2024fzc}. In this context,  the dark matter is a Majorana fermion and its stability is a natural consequence coming from symmetry breaking and the matter content. 
It is well-known that one could observe dark matter signatures in different ways: a) in direct detection experiments one looks for the scattering with matter, b) in collider experiments one looks for missing energy signatures, and c) in indirect detection experiments one can look for gamma lines, neutrino lines and other signatures.
The main goal of this article is to investigate the predictions for gamma lines in the new theory proposed in Ref.~\cite{FileviezPerez:2024fzc}, where a fermionic dark matter candidate is predicted from anomaly cancellation. Since the gamma lines from dark matter annihilation are produced at the quantum level it is very important to have anomaly-free gauge theory. Only in this context it is possible to perform a full study and compute for example higher order processes leading to gamma-ray lines relevant for indirect dark matter searches. For previous studies about gamma lines in gauge theories see Refs.~\cite{Jackson:2009kg,Jackson:2013pjq,Jackson:2013tca,FileviezPerez:2019rcj,Bringmann:2012ez}.

In this article, we investigate in great detail the predictions for gamma lines in the minimal theory for baryon number~\cite{FileviezPerez:2024fzc} taking into account the relic density and direct detection constraints. We perform a full study of the different mechanisms generating gamma lines and discuss the correlation between the gamma lines and the other processes generating a continuum spectrum for photons that could spoil the visibility of the gamma lines. Our main results can be summarized as follows:
\begin{itemize}

\item In this theory the main mechanisms generating gamma lines are: $\chi \chi \to Z_B^* \to \gamma \gamma$ and $\chi \chi \to Z_B^* \to Z \gamma$, where $Z_B$ is the new gauge boson associated to baryon number. 
Here the effective $Z_B \gamma \gamma$ and $Z_B Z \gamma$ couplings are generated at one-loop level, where inside one has the new electrically charged fields needed for anomaly cancellation. 

\item The processes $\chi \chi \to h_B h_B \to \gamma \gamma \gamma \gamma$ and $\chi \chi \to Z_B h_B \to q \bar{q} \gamma \gamma$ produce photons with energy in a relative small region. In this theory the new physical Higgs boson, $h_B$, has a very large decay branching ratio to photons if its mass is below the $WW$ mass threshold and the mixing angle between the Higgses is small, as required by the direct detection experimental bounds.

\item The final state radiation processes $\chi \chi \to Z_B \to q \bar{q} \gamma$ can spoil the visibility of the gamma lines in some scenarios. In this theory these processes occur at tree level but they are velocity suppressed or suppressed by the quark masses.

\item We show that the gamma lines from dark matter annihilation predicted in this theory could be discovered in the near future at gamma-ray telescopes such as CTA.
\end{itemize}

\begin{figure} [h]  
        \centering       
        \includegraphics[width=0.95\textwidth]{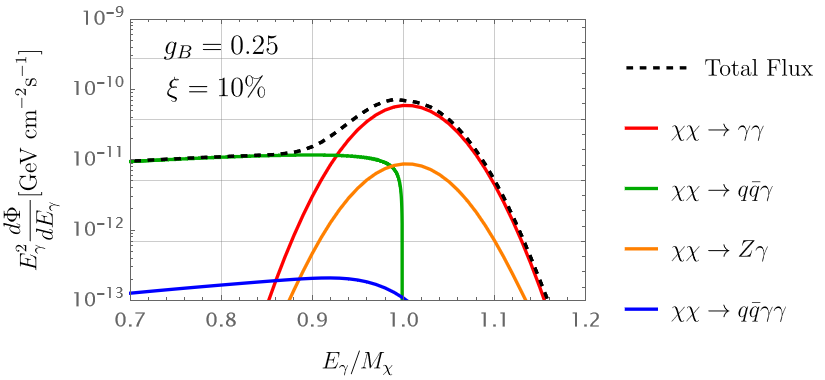}
         \caption{Predictions for gamma lines when we have a $10\%$ energy resolution. As  benchmark scenario we use: $M_\chi=$5.2 TeV, $M_{\ZB}=860$ GeV, $g_B=0.25$, $\sin{\theta_B}=0.01$, $M_{\Psi^-}= 1.2 M_{\chi}$, $M_{\rho^-}=3 M_\chi$ and $M_{h_B}=200$ GeV. For more details see the discussion in Sec.~\ref{gammalines}.}
         \label{Energy1}
\end{figure}

In order to appreciate the importance of our results, we show in Fig.~\ref{Energy1} the predictions for the gamma lines in a benchmark scenario. In red and orange we show the predictions for the gamma lines, $\chi \chi \to \gamma \gamma$ and $\chi \chi \to Z \gamma$, respectively. In green we show the predictions for the final state radiation, $\chi \chi \to Z_B^* \to q \bar{q} \gamma$, while in blue we show the contribution from $\chi \chi \to Z_B h_B \to q \bar{q} \gamma \gamma$. As one can appreciate, one can see clearly the transition from the continuum to the gamma line.
For more details see the details in the next sections.
This article is organized as follows: In Sec.~\ref{theory} we discuss the minimal theory for local baryon number and the dark matter candidates. In Sec.~\ref{dm} we discuss the properties of the dark matter candidate and the constraints from relic density and direct detection experiments. In Sec.~\ref{gammalines} the different mechanisms generating gamma lines are discussed in great detail together with the possible processes such as final state radiation that could spoil the visibility of the gamma lines. We summarize our main results in Sec.~\ref{summary}.
%
\section{THEORY FOR DARK MATTER AND BARYON NUMBER}
\label{theory}
As we have mentioned in the previous section, our main interest is to study the properties of dark matter candidates in a theory where a cold dark matter candidate is predicted by the cancellation of gauge anomalies. This prediction is possible in theories where the total lepton number or baryon number is promoted to a local gauge symmetry. Here we discuss the theory where the baryon number in the SM is promoted to a local gauge symmetry and the theory is based on the gauge group~\cite{FileviezPerez:2024fzc}:
$$SU(3)_C \otimes SU(2)_L \otimes U(1)_Y \otimes U(1)_B$$
In Ref.~\cite{FileviezPerez:2024fzc} it has been pointed out that 
all baryonic anomalies can be cancelled with only four extra fermions with the following quantum numbers:
\begin{eqnarray}
    \Psi_L & \sim & ({\bf{1}}, {\bf{1}},-1,3/4), \
    \Psi_R   \sim  ({\bf{1}},{\bf{1}},-1,-3/4), \nonumber \\
    \chi_L & \sim & ({\bf{1}},{\bf{1}},0,3/4), \ {\text{and}} \
    \rho_L   \sim  ({\bf{1}},{\bf{3}},0,-3/4). \nonumber
    \label{fermions}
\end{eqnarray}
Notice that the field $\chi_L$ is electrically neutral, and the field $\rho_L$ also contains a neutral component. 
One can generate mass for the new fermions if one has the Higgs field, $S \sim ({\bf{1}},{\bf{1}},0,3/2)$, and the following Yukawa interactions:
\begin{eqnarray}
- \mathcal{L} &\supset& \lambda_\rho {\rm{Tr}}(\rho_L^T C \rho_L) S + \lambda_\Psi \bar{\Psi}_L \Psi_R S 
+ \lambda_\chi \chi_L^T C \chi_L S^*  \ + \ {\rm{h.c.}}.
\label{interactions-masses}
\end{eqnarray}
The new electrically charged fields, $\Psi_L$ and $\Psi_R$, can decay through the new interactions between the SM fields and the new fields. This is possible if  we have a new scalar field, $\phi \sim ({\bf{1}},{\bf{1}},0,3/4)$, the interaction term $\lambda_e \bar{\Psi}_L e_R \phi + h.c.$ and the higher-dimensional operators allowed by the gauge symmetries
\footnote{Notice that the dimension five operators: 
\begin{eqnarray*}
- \mathcal{L} &\supset&   y_1 \ell_L^T i \sigma_2 C \rho_L H \phi / \Lambda \ + \ 
 y_2 \ell_L^T  i \sigma_2 C H \chi_L \phi^* / \Lambda \ + \ y_3 H^\dagger \chi_L^T C \rho_L H/\Lambda \ + \ {\rm{h.c.}},
\end{eqnarray*}
are allowed by the gauge symmetry.}.
The scalar potential in this theory is given by 
 \begin{eqnarray}
 V(H,S,\phi)&=&-m_H^2 H ^{\dagger}H+\lambda(H^{\dagger}H)^2-m_s^2 S ^{\dagger}S 
 + \lambda_s (S^{\dagger}S)^2-m_{\phi}^2 \phi ^{\dagger}\phi 
 + \lambda_{\phi}(\phi^{\dagger}\phi)^2 \nonumber \\
 &+& \lambda_1(H^{\dagger}H)S^{\dagger}S + \lambda_2(H^{\dagger}H)\phi^{\dagger}\phi 
 + \lambda_3(S^{\dagger}S)\phi^{\dagger}\phi + \left( \mu S^* \phi \phi + h.c.\right),
 \end{eqnarray}
where $H \sim({\bf{1}},{\bf{2}},1/2,0)$ is the SM Higgs field. 
Here we have the freedom to redefine the scalar fields and work with a positive and real $\mu$-term in the scalar potential. The scalar potential has the global symmetry: $O(4)_H \otimes U(1)$. Notice that when $S \to e^{ia} S$ and $\phi \to e^{ia/2} \phi$, with $a $ being a real constant phase, the potential does not change.
Notice that in this theory the proton is absolutely stable because the baryon number is never broken by one unit.
The scalar fields can be written as 
\begin{eqnarray}
H&=&\begin{pmatrix}
h^+\\
\frac{1}{\sqrt{2}}(v_{0} + h_0) e^{i \sigma_0/v_0} 
\end{pmatrix}, \\
S &=& \frac{1}{\sqrt{2}}\left(v_{S} + h_S \right) e^{i \sigma_s/v_s}, 
\label{S-filed}
\\
\phi &=& \frac{1}{\sqrt{2}} \left( v_\phi+ \phi_R + i \phi_I  \right).
\end{eqnarray}
The kinetic terms for the new Higgses are given by
\begin{equation}
\mathcal{L} \supset (D_\mu S)^\dagger (D^\mu S) +  (D_\mu \phi)^\dagger (D^\mu \phi),
\end{equation}
where 
\begin{eqnarray}
D_\mu S &=& \partial_\mu S + i \frac{3 g_B}{2}  Z_\mu^B S, \  \text{and} \
D_\mu \phi  =  \partial_\mu \phi + i \frac{3 g_B}{4} Z_\mu^B \phi.
\end{eqnarray}
The kinetic terms for the new fermions are given by
\begin{equation}
\mathcal{L} \supset i \bar{\Psi}_L \gamma_\mu D^\mu \Psi_L + i \bar{\Psi}_R \gamma_\mu D^\mu \Psi_R + i \bar{\chi}_L \gamma_\mu D^\mu \chi_L + i {\text{Tr}} \left( \bar{\rho}_L \gamma_\mu D^\mu \rho_L \right),  
\end{equation}
with
\begin{eqnarray}
D^\mu \Psi_L &=& \partial^\mu \Psi_L - i g_1 B^\mu \Psi_L + i \frac{3}{4} g_B Z^\mu_B \Psi_L , \\   
D^\mu \Psi_R &=& \partial^\mu \Psi_R - i g_1 B^\mu \Psi_R - i \frac{3}{4} g_B Z^\mu_B \Psi_R, \\
D^\mu \chi_L &=& \partial^\mu \chi_L + i \frac{3}{4} g_B Z^\mu_B \chi_L, \\ 
D^\mu \rho_L &=& \partial^\mu \rho_L + i g_2 [W^\mu, \rho_L] - i g_B \frac{3}{4}  Z^\mu_B \rho_L.
\end{eqnarray}
The $\rho_L$ field can be written as
\begin{equation}
\rho_L=\frac{1}{\sqrt{2}} 
\left( \begin{matrix}
\rho_L^0 & \sqrt{2} \rho_L^+ \\
\sqrt{2} \rho_L^- & - \rho_L^0
\end{matrix}
\right).
\end{equation}
In the above equations, $B_\mu$ is the gauge field associated to $U(1)_Y$, and $W_\mu$ is the $SU(2)_L$ gauge field which can be written as
\begin{equation}
W_\mu=\frac{1}{2} 
\left( \begin{matrix}
W_\mu^3 & \sqrt{2} W^+_\mu \\
\sqrt{2} W^-_\mu & - W^3_\mu
\end{matrix}
\right).
\end{equation}
In general one can have two main scenarios:
\begin{itemize}
\item $v_\phi=0$: In this case the new charged fermions can decay into the SM charged leptons and the field $\phi$. Here the lightest field between $\phi$, $\rho_L^0$ and $\chi_L$ can be a dark matter candidate because one has the accidental discrete symmetry: 
\begin{equation}
{\mathcal{Z}}_2: \phi \to - \phi, \rho_L \to - \rho_L, \chi_L \to - \chi_L, \Psi_L \to - \Psi_L, \Psi_R \to - \Psi_R,  
\end{equation}
protecting the stability of the lightest field.
Notice that this symmetry is a natural consequence of the gauge symmetry and particle content. This symmetry is also respected by the higher-dimensional operators discussed above.
\item $v_\phi \neq 0$:  
In this case all the new fields needed for anomaly cancellation can decay to the SM fields.
The field $\Psi_L$ can mix with the $e_R$, while the neutral fields, $\rho_L^0$ and $\chi_L$, decay via higher-dimensional operators mentioned above.
\end{itemize}
This theory predicts the following new physical fields (we are interested in the case where $v_\phi=0$):
\begin{itemize}
    \item $Z_B$ is the new gauge boson associated to the local baryon number. The $Z_B$ mass is given by $M_{Z_B}=3 g_B v_S/2$ in the case where the $\phi$ field does not acquire a vacuum expectation value.
    
    \item $h$ is the SM-like Higgs boson defined as: $h= h_0 \cos \theta_B  - h_S \sin \theta_B$.
    \item $h_B$ is the new CP-even Higgs defined as: $h_B=h_0 \sin \theta_B + h_S \cos \theta_B$. 
    \item $\phi$ is a complex scalar field. The masses for $\phi_R$ and $\phi_I$ read as: 
\begin{eqnarray}
M_{\phi_R}^2&=&-m_\phi^2+ \frac{\lambda_2}{2} v_0^2 + \frac{\lambda_3}{2} v_S^2 + \sqrt{2} \mu v_S, \\
M_{\phi_I}^2&=& M_{\phi_R}^2 - 2 \sqrt{2} \mu v_S,
\end{eqnarray}
when $v_\phi=0$.
    \item Two Majorana fermionic fields: $\chi = \chi_L + (\chi_L)^C$ and $\rho^0=\rho_L^0 + (\rho_L^0)^C$ with masses
\begin{eqnarray}
    M_\chi &=&\sqrt{2} \lambda_\chi v_S \ {\text{and}} \
    { M_{\rho^0} = \sqrt{2} \lambda_\rho v_S.}
\end{eqnarray}
    
    \item Two charged fields: $\Psi^-=\Psi_L^- + \Psi_R^-$ and $\rho^-=\rho_L^- + (\rho_L^+)^C$ with masses given by
\begin{eqnarray}
M_{\Psi^-} &=& \frac{1}{\sqrt{2}} \lambda_\Psi v_S \ {\text{and}} \
M_{\rho^-}  =  M_{\rho^0} + \delta M,    
\end{eqnarray}
where the mass splitting $\delta M \approx 166$ MeV is generated at one-loop level~\cite{Cirelli:2005uq}.

\end{itemize}

It is important to mention that the field $\chi_L$ is predicted in the different anomaly-free theories proposed in Refs.~\cite{Duerr:2013dza,FileviezPerez:2014lnj,FileviezPerez:2024fzc} and for this reason we will discuss its properties as a dark matter candidate in the next section. See also the study in Ref.~\cite{Ellis:2017tkh}.
\begin{figure}[h]
         \centering
    \begin{subfigure}[b]{0.45\textwidth}
         \centering
         \includegraphics[width=\textwidth]{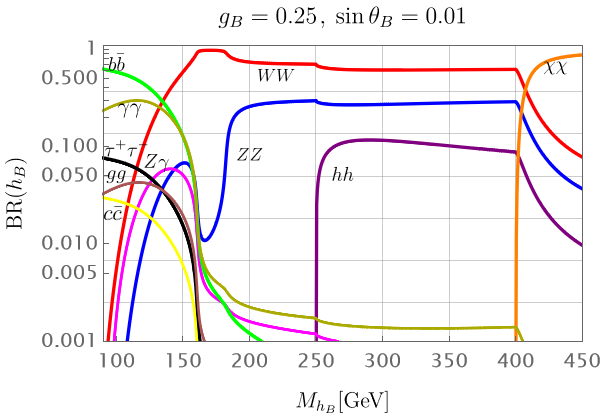}
         \caption{}
    \end{subfigure}  
         \begin{subfigure}[b]{0.45\textwidth}
         \centering
         \includegraphics[width=\textwidth]{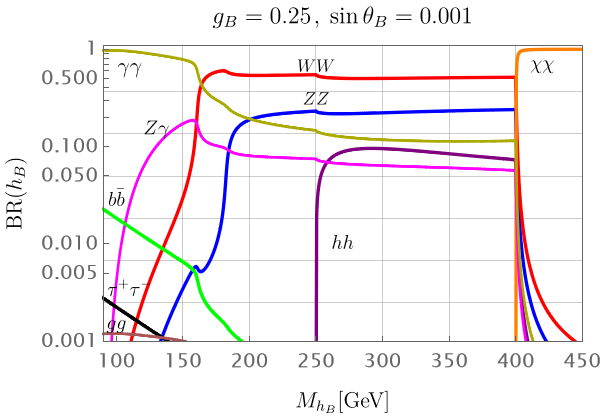}
         \caption{}
    \end{subfigure} 
        \caption{  Branching ratios for the $h_B$ decays. Here we use $M_{\chi}= 200$ GeV, $M_{\ZB}=300$ GeV, $M_{\Psi^-}= 1.2 M_{\chi}$, $M_{\rho^-}=3 M_\chi$, and $g_B=0.25$.}
        \label{Higgs}
\end{figure}

The new Higgs boson, $h_B$, has very interesting properties. In Figs.~\ref{Higgs} we show the branching ratios for the $h_B$ decays in two scenarios. In the first scenario shown in Fig.~\ref{Higgs}a) the gauge coupling $g_B=0.25$ and the mixing angle between the two Higgses is $\sin \theta_B=0.01$. As one can appreciate, the main decays in the low mass region, below the $WW$ mass threshold, are $b \bar{b}$ and $\gamma \gamma$. Notice that the effective $h_B\gamma\gamma$ coupling is generated mainly by the new electrically charged fermions $\rho^-$ and $\Psi^-$, which acquire their masses once $S$ breaks the local $U(1)_B$ gauge symmetry. The decay into the dark matter dominates when it is allowed, in this case we assume $M_\chi=200$ GeV. In Fig~\ref{Higgs}b) we show the same results but in the case when the mixing between the two Higgses is smaller, i.e. $\sin \theta_B=0.001$. In this case the decays into two b-quarks are suppressed and $h_B$ decays mainly into two photons in the low mass region. It is remarkable that this theory predicts a Higgs boson that can have a very large decay branching ratio to photons. We called this type of Higgs ``Cucuyo'' Higgs in a previous study~\cite{Butterworth:2024eyr}. Notice that the properties of the new Higgs are also very important to understand the flux of gamma-ray from dark matter annihilation.

\section{MAJORANA DARK MATTER}
\label{dm}
As we discussed in previous sections, in this model~\cite{FileviezPerez:2024fzc} one can have different dark matter candidates but we focus on the scenario where the Majorana field $\chi$ is the dark matter candidate because this candidate appears in all anomaly-free models based on local baryon number. The properties of $\chi$ as a dark matter candidate has been studied in Refs.~\cite{Duerr:2015vna,FileviezPerez:2019jju,FileviezPerez:2020oxn,Butterworth:2024eyr}.
Here we update the results and include the new LZ bounds for direct detection to achieve our main goal proving a detailed discussion of the gamma lines and the gamma-ray spectrum from dark matter annihilation.

The main annihilation channels for the Majorana dark matter candidate, 
\begin{equation}
\chi=\chi_L + (\chi_L)^C,    
\end{equation}
are
\begin{equation}
 \chi \chi \to \bar{q} q, Z_B Z_B, Z_B h_B, Z_B h, h_B h_B, h h_B, h h, W W, Z Z. \end{equation}
Notice that the annihilation channels, $Z_B h$, $h h_B$, $WW$, and $ZZ$, are suppressed by the small mixing angle $\theta_B$. In this theory, we have only a few free parameters that play a main role in the discussion of our dark matter candidate:
\begin{equation}
g_B, M_\chi, M_{Z_B}, M_{h_B}, \ \text{and} \ \theta_B.
\end{equation}
The Feynman graphs for the annihilation channels are given in Fig.~\ref{annihilation}.
 \begin{figure} [h]  
        \centering   
        \includegraphics[width=0.9\textwidth]{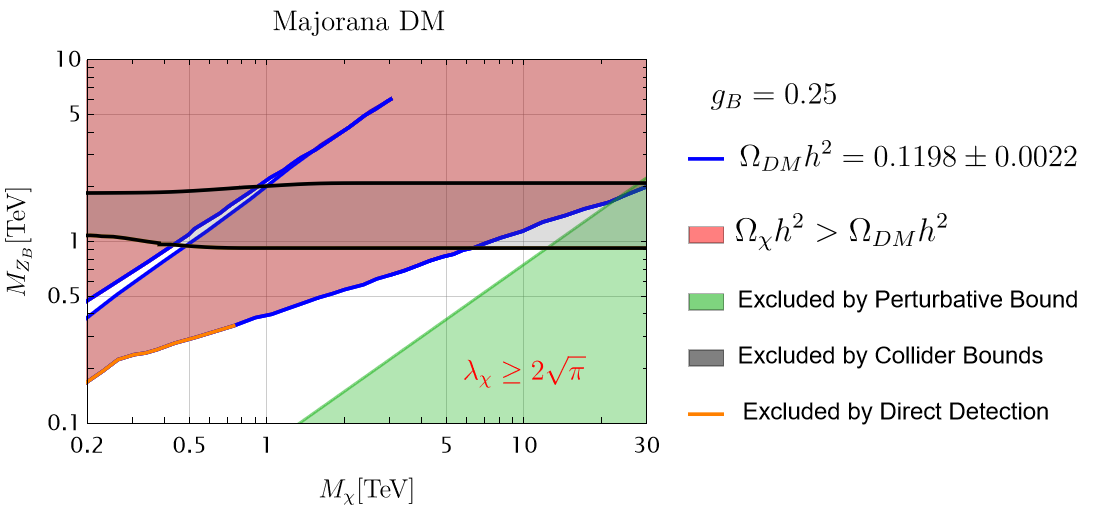}
         \caption{Allowed parameter space in the $M_{Z_B}-M_\chi$ plane when $g_B=0.25$. Here the blue solid line represents $\Omega h^2 = 0.1198\pm 0.0022$ and the red shaded region represents the parameter space when the relic density is greater than 0.12 and the green shaded region is excluded by the perturbative bound on $\lambda_\chi$. For illustration we use $M_{h_B}= 250$ GeV. The region in black is excluded by the current LHC bounds~\cite{Butterworth:2024eyr}, while the orange line is excluded by the LZ direct detection bounds.}
         \label{DM-0.25}
\end{figure}

In Fig.~\ref{DM-0.25} we show the allowed parameter space by the relic density constraints, $\Omega_{\chi} h^2 \leq \Omega_{DM} h^2=0.1198$, 
in the $M_{Z_B}-M_\chi$ plane. The green region is excluded by the perturbative bounds on the Yukawa coupling associated to the dark matter mass, $\lambda_\chi \leq 2 \sqrt{\pi}$, while the orange line is excluded by the LZ direct detection bounds. One can achieve the correct relic density, $\Omega_{DM} h^2=0.1198$, in two main regions: a) in the resonance region, $M_\chi \sim M_{Z_B}/2$, the main annihilation channel is $\chi \chi \to Z_B \to q \bar{q}$ which is velocity suppressed. This region is less generic but tells us what is the maximum allowed value for the gauge boson mass consistent with the relic density constraints. b) In the second region the main annihilation channel is $\chi \chi \to Z_B h_B$. This region is more generic and one can achieve the correct relic density without assuming any peculiar relation between the input parameters. In black we show the excluded region by the current LHC bounds as pointed out in a previous study~\cite{Butterworth:2024eyr}. 
See also Ref.~\cite{Altakach:2024jwk} for the monojet bounds.
In the scenario shown in Fig.~\ref{DM-1.0} the collider bounds are not important because the gauge boson mass is in the multi-TeV region. As in the previous case, we have two main regions allowed by the relic density constraints. Notice that since the gauge coupling is much larger in this case the resonance region is wider and the maximal value for the gauge boson mass is around $15$ TeV. Using these constraints we can study the predictions for gamma lines in the next section.

\begin{figure}
\begin{eqnarray*}
\begin{gathered}
\begin{tikzpicture}[line width=1.5 pt,node distance=1 cm and 1.5 cm]
\coordinate[label = left: $\chi$] (i1);
\coordinate[below right = 1cm of i1](v1);
\coordinate[below left = 1cm of v1, label= left:$n$](i2);
\coordinate[above right = 1cm of v1, label=right: $\chi$] (f1);
\coordinate[below right =  1cm of v1,label=right: $n$] (f2);
\draw[fermionnoarrow] (i1) -- (v1);
\draw[fermion] (i2) -- (v1);
\draw[fermionnoarrow] (v1) -- (f1);
\draw[fermion] (v1) -- (f2);
\draw[fill=gray] (v1) circle (.3cm);
\end{tikzpicture}
\end{gathered} 
&=&
\begin{gathered}
\begin{tikzpicture}[line width=1.5 pt, node distance=1 cm and 1.5 cm, transform shape]
\coordinate[label=left: $\chi$] (i1);
\coordinate[right=1cm of i1] (v1);
\coordinate[below=0.6cm of v1, label=right: {$h_B, h$}] (vaux);
\coordinate[right=1cm of v1, label=right:$\chi$] (f1);
\coordinate[below=1.2 cm of v1] (v2);
\coordinate[left=1 cm of v2, label=left: $n$] (i2);
\coordinate[right=1 cm of v2, label=right: $n$] (f2);
\draw[fermionnoarrow] (i1) -- (v1);
\draw[scalarnoarrow] (v1) -- (v2);
\draw[fermion] (i2) -- (v2); 
\draw[fermion] (v2) -- (f2); 
\draw[fermionnoarrow] (v1) -- (f1);
\draw[fill=black] (v1) circle (.1cm);
\draw[fill=black] (v2) circle (.1cm);
\end{tikzpicture}
\end{gathered}
\quad   \! +
\begin{gathered}
\begin{tikzpicture}[line width=1.5 pt, node distance=1 cm and 1.5 cm,  transform shape]
\coordinate[label=left: $\chi$] (i1);
\coordinate[right=1cm of i1] (v1);
\coordinate[below=0.6cm of v1, label=right: {$Z_B$}] (vaux);
\coordinate[right=1cm of v1, label=right:$\chi$] (f1);
\coordinate[below=1.2 cm of v1] (v2);
\coordinate[left=1 cm of v2, label=left: $n$] (i2);
\coordinate[right=1 cm of v2, label=right: $n$] (f2);
\draw[fermionnoarrow] (i1) -- (v1);
\draw[vector] (v1) -- (v2);
\draw[fermion] (i2) -- (v2); 
\draw[fermion] (v2) -- (f2); 
\draw[fermionnoarrow] (v1) -- (f1);
\draw[fill=cyan] (v1) circle (.1cm);
\draw[fill=cyan] (v2) circle (.1cm);
\end{tikzpicture}
\end{gathered}
\end{eqnarray*}
\caption{Feynman graphs for the dark matter-nucleon effective interaction.}
\label{graphs-DD}
\end{figure}
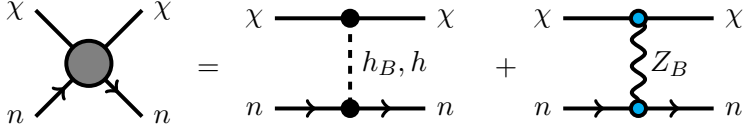

The dark matter-nucleon interactions are mediated by the SM Higgs, the new Higgs $h_B$ and the new gauge boson $Z_B$, see Fig.~\ref{graphs-DD}.
The nucleon-DM elastic spin-independent cross section mediated by the $Z_B$ gauge boson is given by
\begin{eqnarray}
\sigma_{\chi N}^\text{SI}(Z_B)&=& \frac{27}{32 \pi}\frac{g_B^4 M_N^2}{M_{Z_B}^4} v^2,
\end{eqnarray}
while the Higgs contributions read as
\begin{eqnarray}
\sigma_{\chi N}^\text{SI}(h_i)&=&\frac{9 G_F}{2\sqrt{2}  \pi}\sin^2 \theta_B \cos^2 \theta_B M_N^4 \frac{g_B^2 M_\chi^2}{M_{Z_B}^2}\left(\frac{1}{M_{h}^2}-\frac{1}{M_{h_B}^2}\right)^2 f_N^2,
\end{eqnarray}
$M_N$ corresponds to the nucleon mass, $G_F$ is the Fermi constant  and for the effective Higgs-nucleon-nucleon coupling we take $f_N=0.3$ \cite{Hoferichter:2017olk}. The axial coupling between $\chi$ and $Z_B$ leads to velocity suppression of the cross-section and we can write,
\begin{equation}
\sigma_{\chi N}^\text{TOT} = \sigma_{\chi N}(h_i) + \sigma_{\chi N}^\text{0}(Z_B) v^2.
\end{equation}
The bounds coming from direct detection experiments are obtained under the assumption that the leading order in the cross-section is velocity independent. In order to apply these bounds we proceed as follows,
\begin{equation}
 \sigma_{\chi N}(h_i) + \sigma^0_{\chi N}(Z_B) v_\text{eff}^2 \leq \sigma_{\chi N}^{\text{DDexp}},
\end{equation}
where  $\sigma_{\chi N}^{\text{DDexp}}$ is the upper bound on the scattering cross-section given by the direct detection experiments, and the effective velocity is given by the ratio $\overline{v^3}/\overline{v}$, where the average velocity is the velocity of the dark matter convoluted with a Maxwell-Boltzmann distribution. Here the effective velocity is $v_\text{eff} \approx 0.001 \, c $. 
Notice that the contribution to the spin-dependent cross-section is suppressed by the momentum transferred
and since in our case we have the elastic scattering limit this contribution is highly-suppressed.
\begin{figure} [h]  
        \centering       \includegraphics[width=0.85\textwidth]{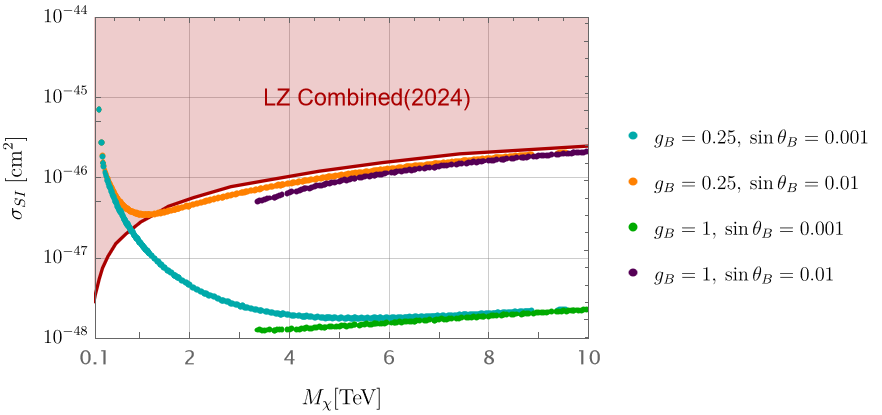}
         \caption{Dark matter-nucleon spin independent cross section when $M_{h_B}=200$ GeV. The LZ bounds~\cite{LZ} are excluding the region in red.}
         \label{DD}
\end{figure}

In Fig.~\ref{DD} we show the predictions for the direct detection nucleon-dark matter cross section for different scenarios where one achieves the correct value for the relic density. The region in red is excluded by the new LZ direct detection bounds. We show the four scenarios: a) $g_B=0.25$ and $\sin \theta_B=0.001$ b) $g_B=0.25$ and $\sin \theta_B=0.01$ 
c) $g_B=1$ and $\sin \theta_B=0.001$ and d) $g_B=1$ and $\sin \theta_B=0.01$. As one can appreciate, some scenarios with $\sin \theta_B=0.01$ are ruled out by the LZ bounds, and only a small region when $g_B=0.25$ is excluded. This results justify our choice for the value of the mixing angle used in the previous section to study the Higgs decays which is very important to predict the gamma-ray spectrum coming from dark matter annihilation.
%
\section{GAMMA LINES}
\label{gammalines}
One of the most striking signatures in dark matter models is related to the possibility to predict gamma lines from dark matter annihilation.
In this section, we study carefully the predictions for gamma lines taking into account the relic density and direct detection constraints. Here we present the numerical results for the annihilation cross sections and the flux for gamma lines. 

In this theory the new Higgs $h_B$ can be light and have a large coupling to the dark matter candidate. Therefore, one can expect a Sommerfeld enhancement that is relevant to predict the annihilation cross sections.
\begin{figure}[h]  
        \centering 
        \includegraphics[width=0.60\textwidth]{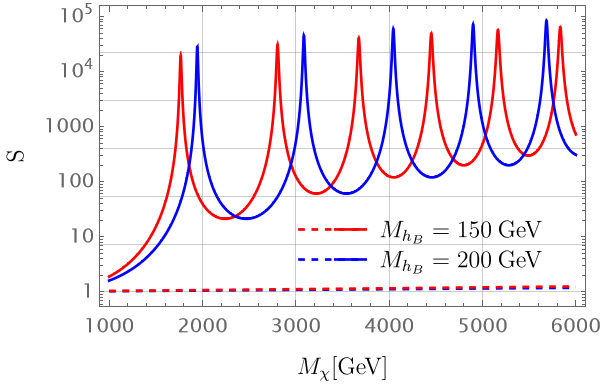}
         \caption{Sommerfeld enhancement mediated by the new Higgs boson as a function of dark matter mass. Here we use $g_B =0.25$ and $v=10^{-3}$. The red (blue) solid line shows the result for $M_{h_B}=150 (200)$ GeV when $M_{\ZB}=500$ GeV. The dashed lines shows the Sommerfeld Enhancement when $M_{Z_B}= 2 M_{\chi}$ which is the resonance region.}
         \label{Som1}
\end{figure}
The Sommerfeld enhancement for the Yukawa potential can be obtained by approximating the Hulthen Potential and the resulting analytical approximation gives~\cite{Cassel:2009wt}
 \begin{equation}
     S = \frac{\pi}{\epsilon_{\nu}}\frac{\sinh{\left(\frac{2 \pi \epsilon_\nu}{\pi^2 \epsilon_\phi/6}\right)}}{\cosh{\left(\frac{2 \pi \epsilon_\nu}{\pi^2 \epsilon_\phi/6}\right)} - \cos{\left(2 \pi \sqrt{\frac{1}{\pi^2 \epsilon_\phi/6}-\frac{\epsilon_\nu^2}{(\pi^2 \epsilon_\phi/6)^2}} \right)}}
 \end{equation}
 with
 \begin{equation}
 \epsilon_\nu = v/\alpha_\chi, \ \epsilon_\phi = M_{h_B}/(\alpha_\chi M_{\chi}), \ \alpha_\chi = \frac{1}{4 \pi}\left(\frac{3g_B M_\chi}{2 M_{\ZB}}\right)^2,
 \end{equation}
 $v$ is the dark matter velocity in the center of mass frame and $M_{h_B}$ is the mediator (Higgs) mass. See for example Refs.~\cite{Hisano:2004ds,Feng:2010zp} for the impact of Sommerfeld enhancement in dark matter models. 
 
 In Fig.~\ref{Som1} we show the values of the Sommerfeld enhancement when we use $g_B =0.25$ and $v=10^{-3}$. In red and blue we show the results for $M_{h_B}=150$ GeV and $M_{h_B}=200$ GeV, respectively. Clearly, this enhancement factor is key to predict the gamma lines and see if one can hope to test the predictions at current or future experiments. Now, we are ready to discuss the different mechanisms giving rise to gamma-lines and the processes such as the final state radiation that could spoil the visibility of gamma lines.

In this theory there two types of gamma lines, $\chi \chi \to \gamma \gamma$ and $\chi \chi \to Z \gamma$, and two processes where the energy of the photons are in a range, $\chi \chi \to h_B h_B \to \gamma \gamma \gamma \gamma$ and $\chi \chi \to Z_B h_B \to q \bar{q} \gamma \gamma$. The final state radiation processes $\chi \chi \to q \bar{q} \gamma$ are also very important to understand the visibility of the gamma lines. Here we provide a detailed discussion of each of these processes: 
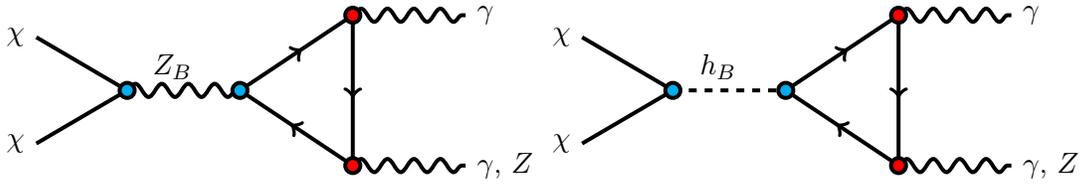
\begin{figure}
\begin{eqnarray*}    
\begin{gathered}
\begin{tikzpicture}[line width=1.5 pt,node distance=1 cm and 1.5 cm]
\coordinate[label =left: $\chi$] (i1);
\coordinate[below right= 1cm of i1](p1);
\coordinate[below left = 1cm of p1, label= left:$\chi$](i2);
\coordinate[right=0.5 cm of p1](j1);
\coordinate[right=0.6 cm of j1, label=$Z_B$](vaux);
\coordinate[right = of j1](p2);
\coordinate[right=0.55 cm of p2](vmare);
\coordinate[above=0.5cm of vmare](vaux1);
\coordinate[below=1.1cm of vmare](vaux2);
\coordinate[right= 1cm of vmare](vaux3);
\coordinate[above right = of p2](v1a);
\coordinate[right = of v1a, label=right: $\gamma$] (v1);
\coordinate[below right = of p2](v2a);
\coordinate[right = of v2a,label=right: $\gamma \text{, }Z$] (v2);
\draw[fermionnoarrow] (i1) -- (j1);
\draw[fermionnoarrow] (i2) -- (j1);
\draw[vector] (j1) -- (p2);
\draw[vector] (v1a) -- (v1);
\draw[fermion] (v1a)--(v2a);
\draw[fermion](p2)--(v1a);
\draw[fermion] (v2a)--(p2);
\draw[vector] (v2a) -- (v2);
\draw[fill=cyan] (j1) circle (.1cm);
\draw[fill=cyan] (p2) circle (.1cm);
\draw[fill=red] (v1a) circle (.1cm);
\draw[fill=red] (v2a) circle (.1cm);
\end{tikzpicture}
\end{gathered}
\begin{gathered}
\begin{tikzpicture}[line width=1.5 pt,node distance=1 cm and 1.5 cm]
\coordinate[label =left: $\chi$] (i1);
\coordinate[below right= 1cm of i1](p1);
\coordinate[below left = 1cm of p1, label= left:$\chi$](i2);
\coordinate[right=0.5 cm of p1](j1);
\coordinate[right=0.6 cm of j1, label=$h_B$](vaux);
\coordinate[right = of j1](p2);
\coordinate[right=0.55 cm of p2](vmare);
\coordinate[above=0.5cm of vmare](vaux1);
\coordinate[below=1.1cm of vmare](vaux2);
\coordinate[right= 1cm of vmare](vaux3);
\coordinate[above right = of p2](v1a);
\coordinate[right = of v1a, label=right: $\gamma$] (v1);
\coordinate[below right = of p2](v2a);
\coordinate[right = of v2a,label=right: $\gamma \text{, }Z$] (v2);
\draw[fermionnoarrow] (i1) -- (j1);
\draw[fermionnoarrow] (i2) -- (j1);
\draw[scalarnoarrow] (j1) -- (p2);
\draw[vector] (v1a) -- (v1);
\draw[fermion] (v1a)--(v2a);
\draw[fermion](p2)--(v1a);
\draw[fermion] (v2a)--(p2);
\draw[vector] (v2a) -- (v2);
\draw[fill=cyan] (j1) circle (.1cm);
\draw[fill=cyan] (p2) circle (.1cm);
\draw[fill=red] (v1a) circle (.1cm);
\draw[fill=red] (v2a) circle (.1cm);
\end{tikzpicture}
\end{gathered}
\end{eqnarray*}
\caption{Feynman graphs for the gamma lines from dark matter annihilation.}
\label{FGlines}
\end{figure}
\begin{itemize}
\item $\chi \chi \to Z_B^*, h_B^* \to \gamma \gamma$: 

In this case the energy of the gamma line is $E_\gamma = M_\chi$ and the annihilation cross section is given by
\begin{equation}
 \sigma v ({{\chi}\chi \to \gamma \gamma}) =\frac{\alpha^2}{\pi^3} \frac{g_B^4 n_\chi^2 M_\chi^2}{M_{Z_B}^4}\frac{(4M_\chi^2 - M_{Z_B}^2)^2}{(4M_\chi^2-M_{Z_B}^2)^2+\Gamma_{Z_B}^2M_{Z_B}^2} \left | \sum_{f_+} N_c^{F} n_A^{F}Q_{F}^2M_{F}^2 C_0^\gamma \right |^2,
\label{ACgg}
\end{equation}
Here we have used the relevant interactions using the following convention:
\begin{equation}
{\cal L} \supset g_B n_A^{F} \ \overline{F} \gamma^\mu \gamma^5 F Z^B_\mu, 
\end{equation}
where $F$ are the charged fermions entering in the loops, 
$n_A^{\Psi^-}= 3/4$ and $n_A^{\rho^-}= -3/4$. 
The loop function
\begin{eqnarray}
    C_0^\gamma &=&\frac{1}{2s} ln^2 \left( \frac{\sqrt{1 - 4 M_F^2/s}-1}{\sqrt{1 - 4 M_F^2/s}+1} \right).
\end{eqnarray}
is evaluated at $s=4 M_\chi^2$. See Figs.~\ref{FGlines} and~\ref{Effective-couplings} for the relevant Feynman graphs. Here the contribution mediated by the new Higgs is velocity suppressed.
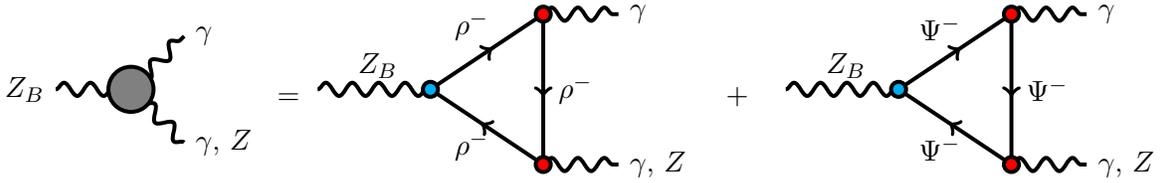
\begin{figure}
\begin{eqnarray*}
\begin{gathered}
\begin{tikzpicture}[line width=1.5 pt,node distance=1 cm and 1.5 cm]
\coordinate[label = left: $Z_B$] (i1);
\coordinate[right = 1cm of i1](v1);
\coordinate[above right = 1cm of v1,label=right: $\gamma$] (f1);
\coordinate[below right =  1cm of v1, label=right: $\gamma \text{, }Z$] (f2);
\draw[vector] (i1) -- (v1);
\draw[vector] (v1) -- (f1);
\draw[vector] (f2) -- (v1);
\draw[fill=gray] (v1) circle (.3cm);
\end{tikzpicture}
\end{gathered} 
&=&
\begin{gathered}
\begin{tikzpicture}[line width=1.5 pt,node distance=1 cm and 1.5 cm]
\coordinate[](j1);
\coordinate[right = of j1](p2);
\coordinate[right=0.55 cm of p2](vmare);
\coordinate[above=0.5cm of vmare,label=$\rho^-$](vaux1);
\coordinate[below=1.1cm of vmare,label=$\rho^-$](vaux2);
\coordinate[right= 1cm of vmare,label=right:$\rho^-$](vaux3);
\coordinate[right = 0.8cm of j1, label=$Z_B$];
\coordinate[above right = of p2](v1a);
\coordinate[right = 1cm of v1a, label=right: $\gamma$] (v1);
\coordinate[below right = of p2](v2a);
\coordinate[right = 1 cm of v2a, label=right: $\gamma \text{, }Z$] (v2);
\draw[vector] (j1) -- (p2);
\draw[vector] (v1a) -- (v1);
\draw[fermion] (v1a)--(v2a);
\draw[fermion](p2)--(v1a);
\draw[fermion] (v2a)--(p2);
\draw[vector] (v2a) -- (v2);
\draw[fill=red] (v1a) circle (.1cm);
\draw[fill=red] (v2a) circle (.1cm);
\draw[fill=cyan] (p2) circle (.1cm);
\end{tikzpicture}
\end{gathered}
\quad   \! + \quad
\begin{gathered}
\begin{tikzpicture}[line width=1.5 pt,node distance=1 cm and 1.5 cm]
\coordinate[](j1);
\coordinate[right = of j1](p2);
\coordinate[right=0.55 cm of p2](vmare);
\coordinate[above=0.5cm of vmare,label=$\Psi^-$](vaux1);
\coordinate[below=1.1cm of vmare,label=$\Psi^-$](vaux2);
\coordinate[right= 1cm of vmare,label=right:$\Psi^-$](vaux3);
\coordinate[right = 0.8cm of j1, label=$Z_B$];
\coordinate[above right = of p2](v1a);
\coordinate[right = 1cm of v1a, label=right: $\gamma$] (v1);
\coordinate[below right = of p2](v2a);
\coordinate[right = 1 cm of v2a, label=right: $\gamma \text{, }Z$] (v2);
\draw[vector] (j1) -- (p2);
\draw[vector] (v1a) -- (v1);
\draw[fermion] (v1a)--(v2a);
\draw[fermion](p2)--(v1a);
\draw[fermion] (v2a)--(p2);
\draw[vector] (v2a) -- (v2);
\draw[fill=red] (v1a) circle (.1cm);
\draw[fill=red] (v2a) circle (.1cm);
\draw[fill=cyan] (p2) circle (.1cm);
\end{tikzpicture}
\end{gathered}
\end{eqnarray*}
\caption{Effective $Z_B \gamma \gamma$ and $Z_B \gamma Z$ couplings generated at one-loop level.}
\label{Effective-couplings}
\end{figure}
\begin{figure} [h]  
        \centering 
        \includegraphics[width=0.90\textwidth]{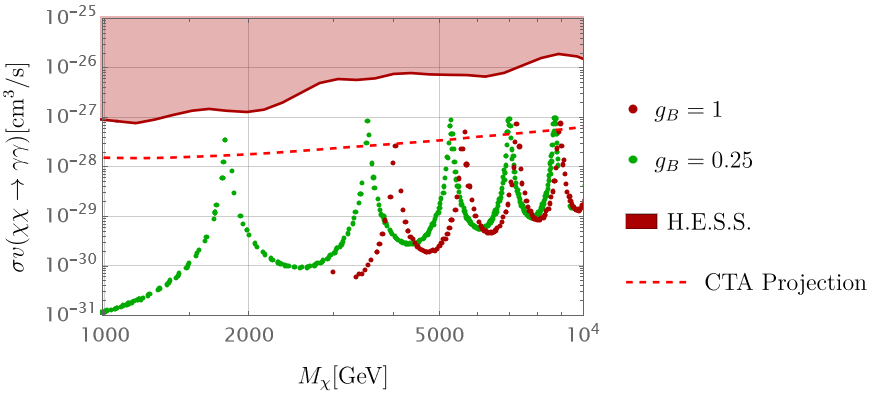}
         \caption{Annihilation cross section $\sigma v (\chi \chi \rightarrow \gamma\gamma)$ as a function of $M_{\chi}$ for different values of $g_B$. Here we used the corresponding values of $M_{Z_B}$ that satisfy the relic density constraints.  Here we used $M_{h_B}$=150 GeV, $M_{\Psi^-}= 1.2 M_\chi$ and $M_{\rho^-}=3 M_{\chi}$. The H.E.S.S bounds are shown in red, while the dashed red line shows the CTA~\cite{Abe_2024} projected bounds. }
         \label{gammaline1}
\end{figure}

In Fig.~\ref{gammaline1} we show the predictions for the cross section corresponding to the gamma line, $\chi \chi \to \gamma \gamma$, for different values of the gauge couplings. In red we show the bounds from the H.E.S.S collaboration and the red dashed line shows the projected bounds from the CTA collaboration~\cite{Abe_2024}. The green (red) dots correspond to the predictions for $g_B=0.25$ ($g_B=1.0$). It is interesting that a few scenarios are close to the CTA protected bounds. In Fig.~\ref{gammaline1} each point corresponds to scenarios where one achieves the correct relic density. These results tell us that there is a hope to test this theory for dark matter in the near future.

The flux for the gamma lines is given by
\begin{equation}
\frac{d\Phi_{\gamma \gamma}}{dE_\gamma}= \frac{n_\gamma }{8 \pi M_\chi^2} \frac{ d ( \sigma v_\text{rel} (\chi \chi \to \gamma \gamma))}{dE_\gamma}J_\text{ann}
=\frac{n_\gamma ( \sigma v_\text{rel} (\chi \chi \to \gamma \gamma) )}{8\pi M_\chi^2}\frac{dN_{\gamma \gamma}}{dE_\gamma} J_\text{ann},
\end{equation}
Here $n_\gamma=2$ and the spectrum function is given by
\begin{equation}
\frac{dN_{\gamma \gamma}}{dE_\gamma}=\int_0^\infty dE_0 \,  \delta(E_0 - M_\chi) \, G(E_\gamma, \xi /\omega, E_0),
\end{equation}
Here, we use a Gaussian function to model the detector resolution, $G(E_\gamma,\xi/\omega, E_0)$, which reads as
\begin{equation}
G(E_\gamma,\xi/\omega, E_0)=\frac{1}{\sqrt{2\pi}E_0(\xi/\omega)} e^{ -\frac{(E_\gamma - E_0)^2}{2E_0^2(\xi/\omega)^2}},
\end{equation}
where $\xi$ is the energy resolution and  $\omega = 2 \sqrt{2 \rm{log} 2} \approx 2.35$ determines the full width at half maximum, with the standard deviation given by $\sigma_0=E_0 \xi/w$.

In Fig.~\ref{fluxgammagamma} we show the predictions for the flux of photons produced in $\chi \chi \to \gamma \gamma$ in two different scenarios: 1) Scenario I: $g_B=0.25$, $M_{h_B}=150$ GeV, $M_\chi=5.2$ TeV, and $M_{Z_B}=860$ GeV, 2) Scenario II: $g_B=1.0$, $M_{h_B}=150$ GeV, $M_\chi=4150$ GeV, and $M_{Z_B}=3.5$ TeV. In both scenarios, we assume $M_{\Psi^-}= 1.2 M_\chi$ and $M_{\rho^-}=3 M_{\chi}$ for the masses of the new electrically charged fermions needed for anomaly cancellation. In both cases we used $J_{ann}=13.9 \times 10^{22}$ GeV$^2$ cm$^{-5}$ for the numerical predictions. 
\begin{figure}[h]
         \centering
    \begin{subfigure}[b]{0.45\textwidth}
         \centering
         \includegraphics[width=\textwidth]{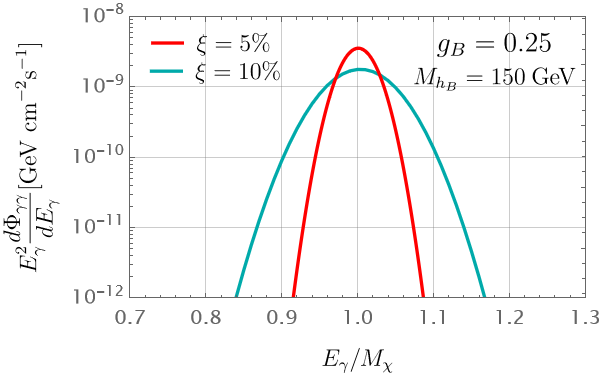}
         \caption{}
    \end{subfigure}  
         \begin{subfigure}[b]{0.45\textwidth}
         \centering
         \includegraphics[width=\textwidth]{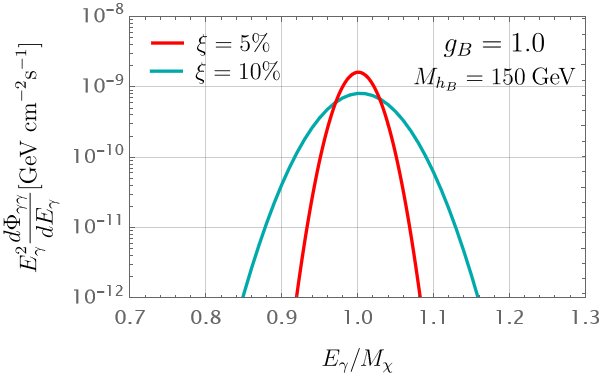}
         \caption{}
    \end{subfigure} 
        \caption{Photon flux from $\chi \chi \rightarrow \gamma \gamma $ as a function of $E_\gamma /M_\chi$ for different energy resolutions. In Fig (a) one shows the results in scenario I, while Fig(b) shows the flux in scenario II. }
        \label{fluxgammagamma}
\end{figure}
%
\item $\chi \chi \to Z_B^*, h_B^* \to \gamma Z$: 

One can have a second gamma-line from the $\chi \chi \to Z \gamma$ annihilation. 
The energy of the gamma lines in this case is given by
\begin{equation}
E_\gamma = M_\chi \left(1 - \frac{M_Z^2}{4 M_{\chi}^2} \right),
\label{Eq1}
\end{equation}
and the annihilation cross section reads as
\begin{equation}
\sigma v (\chi \chi \to \gamma Z) =\frac{\alpha^2 \, g_B^4 n_\chi^2 }{32 \pi^3 \sin^2 2\theta_W}\frac{(4M_\chi^2-M_Z^2)^3}{ (4M_\chi^2-M_{Z_B}^2)^2+\Gamma_{Z_B}^2 M_{Z_B}^2}\frac{(M_{Z_B}^2-4M_\chi^2)^2}{M_\chi^4M_{Z_B}^4} \left| \sum_f N_c^{F} Q_{F}  n_A^{F} g_{V}^{F} 2M_{F}^2 C_0^Z \right|^2,
\label{ACgZ}
\end{equation}
In this case the loop function is given by
\begin{eqnarray}
    C_0^Z &=&\frac{1}{2(M_Z^2 -s)} \left( ln^2 \left( \frac{\sqrt{1 - 4 M_F^2/M_Z^2}-1}{\sqrt{1 - 4 M_F^2/M_Z^2}+1} \right) - ln^2 \left( \frac{\sqrt{1 - 4 M_F^2/s}-1}{\sqrt{1 - 4 M_F^2/s}+1} \right) \right).
\end{eqnarray}
The new electrically charged Higgses inside the loop couple to the $Z$ gauge boson as follows:
\begin{equation}
{\cal L} \supset  - \frac{e}{\sin \theta_W \cos \theta_W} g_V^F \overline{F} \gamma^\mu  F Z_\mu,
\end{equation}
with $g_V^{\Psi^-}=\sin^2{\theta_W}$  and ${g_V^{\rho^-}=-\cos^2{\theta_W}}$. 

In Fig.~\ref{gammaline2} we show the predictions for $\sigma v(\chi \chi \rightarrow Z\gamma) $ as a function of $M_{\chi}$ for different values of $g_B$. We follow the same convention and color code as in Fig.~\ref{gammaline1}. Notice that the values for this cross section are larger than in the previous case because there is no cancellation between the contributions of the two new fermions inside the loop. As one can appreciate, one could test the predictions in most of the scenarios in the near future at CTA. 
\begin{figure} [h]  
        \centering       
        \includegraphics[width=0.85\textwidth]{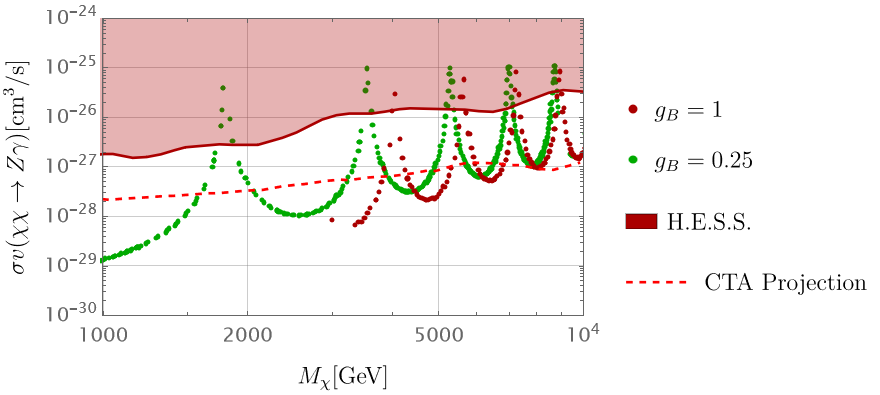}
         \caption{Values of $\sigma v(\chi \chi \rightarrow Z\gamma) $ as a function of $M_{\chi}$ for different values of $g_B$. Here we used $M_{h_B}$=150 GeV, $M_{\Psi^-}= 1.2 M_\chi$ and $M_{\rho^-}=3 M_{\chi}$.}
         \label{gammaline2}
\end{figure}

The flux for the gamma lines for this annihilation channel is given by
\begin{equation}
\frac{d\Phi_{\gamma Z}}{dE_\gamma}= \frac{n_\gamma }{8 \pi M_\chi^2} \frac{ d ( \sigma v_\text{rel} (\chi \chi \to \gamma Z))}{dE_\gamma}J_\text{ann}
=\frac{n_\gamma ( \sigma v_\text{rel} (\chi \chi \to \gamma Z))}{8\pi M_\chi^2}\frac{dN_{\gamma Z}}{dE_\gamma} J_\text{ann},
\end{equation}
Here $n_\gamma=1$ and
\begin{equation}
\frac{dN_{\gamma Z}}{dE_\gamma}=\int_0^\infty dE_0 \,  W_{\gamma Z} \, G(E_\gamma, \xi /\omega, E_0),
\label{spectral}
\end{equation}
where 
\begin{equation}
W_{\gamma Z} = \frac{1}{\pi}\frac{4M_\chi M_Z \Gamma_Z}{(4M_\chi^2-4M_\chi E_0 - M_Z^2)^2+ \Gamma_{Z}^2M_Z^2}.
\label{BWS}
\end{equation}
Here $M_Z$ and $\Gamma_Z$ are the mass and the decay width for the $Z$ gauge boson, respectively. See Ref.~\cite{Duerr:2015aka} for details related to calculation of the spectral function for gamma lines in dark matter models. 

In Fig.~\ref{fluxZgamma} the predictions for the flux of the gamma line from $\chi \chi \to \gamma Z$ is shown. The predictions are shown for scenarios I and II as in the previous case.
Notice that the flux is much smaller than in the case discussed above even if the annihilation cross section $\sigma v(\chi \chi \rightarrow Z\gamma) \gg \sigma v(\chi \chi \rightarrow \gamma \gamma)$. The spectral function in Eq.~(\ref{spectral}) is suppressed due to the 3-body phase space as discussed in Ref.~\cite{Duerr:2015aka}.
\begin{figure}[h]
         \centering
    \begin{subfigure}[b]{0.45\textwidth}
         \centering
         \includegraphics[width=\textwidth]{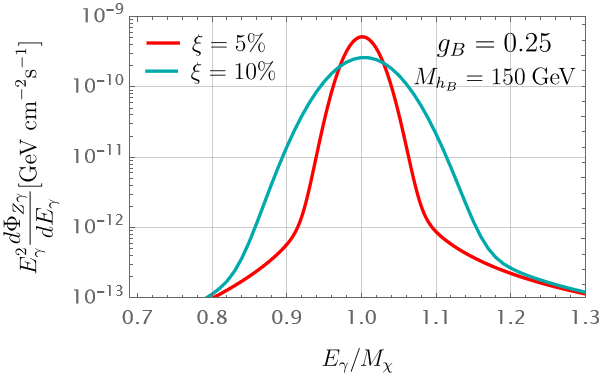}
         \caption{}
    \end{subfigure}  
         \begin{subfigure}[b]{0.45\textwidth}
         \centering
         \includegraphics[width=\textwidth]{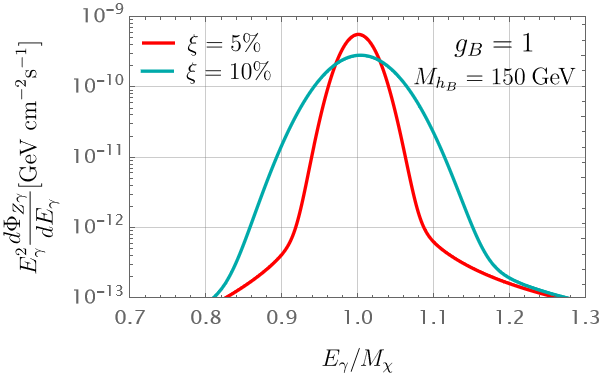}
         \caption{}
    \end{subfigure} 
        \caption{Photon flux from $\chi \chi \rightarrow Z \gamma $ as a function of $E_\gamma /M_\chi$ and different energy resolutions. In the left-panel we show the energy spectrum in scenario I, while in the right-panel one has results for scenario II.}
        \label{fluxZgamma}
\end{figure}
%
\item $\chi \chi \to h_B h_B \to \gamma \gamma \gamma \gamma$:
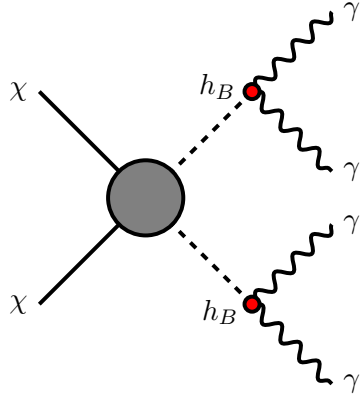
\begin{figure}
\begin{eqnarray*}    
\begin{gathered}
\begin{tikzpicture}[line width=1.5 pt,node distance=1 cm and 1.5 cm]
\coordinate[label = left: $\chi$] (i1);
\coordinate[below right = 2cm of i1](v1);
\coordinate[below left = 2cm of v1, label= left:$\chi$](i2);
\coordinate[above right = 2cm of v1] (f1);
\coordinate[above left = 0.1 cm of f1, label=left: $h_B$];
\coordinate[below right =  2cm of v1] (f2);
\coordinate[below left = 0.6 cm of f2, label=$h_B$];
\coordinate[above right =  1.5cm of f2,label=right: $\gamma$] (f3);
\coordinate[below right =  1.5cm of f2,label=right: $\gamma$] (f4);
\coordinate[above right =  2cm of v1] (g2);
\coordinate[above right =  1.5cm of g2,label=right: $\gamma$] (g3);
\coordinate[below right =  1.5cm of g2,label=right: $\gamma$] (g4);
\draw[fermionnoarrow] (i1) -- (v1);
\draw[fermionnoarrow] (i2) -- (v1);
\draw[scalarnoarrow] (v1) -- (f1);
\draw[scalarnoarrow] (f2) -- (v1);
\draw[vector] (f3) -- (f2);
\draw[vector] (f2) -- (f4);
\draw[vector] (g3) -- (g2);
\draw[vector] (g2) -- (g4);
\draw[fill=gray] (v1) circle (.5cm);
\draw[fill=red] (f1) circle (.1cm);
\draw[fill=red] (f2) circle (.1cm);
\end{tikzpicture}
\end{gathered} 
\end{eqnarray*}
\caption{Graph for $\chi \chi \rightarrow h_B h_B \rightarrow 4\gamma $.}
\label{hBdecays}
\end{figure}

In the previous section we have shown that the ``Cucuyo'' Higgs $h_B$ can have a very large decay branching ratio into two photons. Therefore, in any dark matter annihilation process where we have the $h_B$ Higgs, one has extra contributions to the photon energy spectrum. In this case the energy of the photons from this annihilation channel is in the range~\cite{Ibarra:2015tya}:
\begin{equation}
\frac{M_\chi}{2} \left( 1 - \sqrt{1- \frac{M_{h_B}^2}{M_\chi^2}}\right) \leq E_\gamma \leq  \frac{M_\chi}{2} \left( 1 + \sqrt{1 - \frac{M_{h_B}^2}{M_\chi^2}}\right).  
\end{equation}
\begin{figure}[h]
         \centering
         \includegraphics[width=0.65\textwidth]{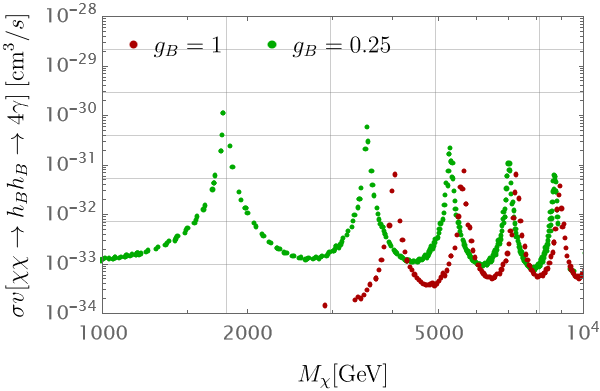}
         \caption{Predictions for $\sigma v (\chi \chi \to h_B h_B)) \times \text{BR}^2(h_B \to 2 \gamma)$ when $\sin{\theta_B}=0.01$ and $M_{h_B}=150$ GeV.}
         \label{crosssectionhBhB}
\end{figure}
See Fig.~\ref{hBdecays} for the relevant graph.

In Fig.~\ref{crosssectionhBhB} we show the numerical predictions for the cross section, $\sigma v (\chi \chi \to h_B h_B)) \times \text{BR}^2(h_B \to 2 \gamma)$ when $\sin{\theta_B}=0.01$ and $M_{h_B}=150$ GeV.  
In this scenario the decay branching ratio into two photons is large but the annihilation cross section, $\chi \chi \to h_B h_B$, is velocity suppressed. This is the main reason why the cross section shown in Fig.~\ref{crosssectionhBhB} is much smaller than in the previous channels discussed above.

The flux for the photons from this annihilation channel is given by
\begin{equation}
\frac{d\Phi_{4 \gamma }}{dE_\gamma}
=\frac{  4 ( \sigma v_\text{rel} (\chi \chi \to h_B h_B)) \times BR^2(h_B \to 2 \gamma)}{8\pi M_\chi^2}\frac{dN_{4\gamma}}{dE_\gamma} J_\text{ann},
\end{equation}
where
\begin{equation}
\frac{dN_{4 \gamma}}{dE_\gamma}=\int_0^\infty dE_0 \,  W_{4\gamma} \, G(E_\gamma, \xi /\omega, E_0),
\end{equation}
with
\begin{equation}
W_{4\gamma} = \frac{1}{(E^{max}_\gamma-E_\gamma^{min})}\Theta (E_\gamma - E_\gamma^{min}) \Theta (E_\gamma^{max} - E_\gamma).
\end{equation}
\begin{figure}[h]
         \centering
    \begin{subfigure}[b]{0.45\textwidth}
         \centering
         \includegraphics[width=\textwidth]{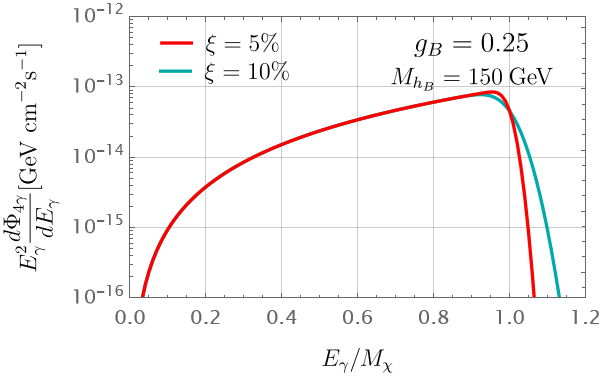}
         \caption{}
    \end{subfigure}  
         \begin{subfigure}[b]{0.45\textwidth}
         \centering
         \includegraphics[width=\textwidth]{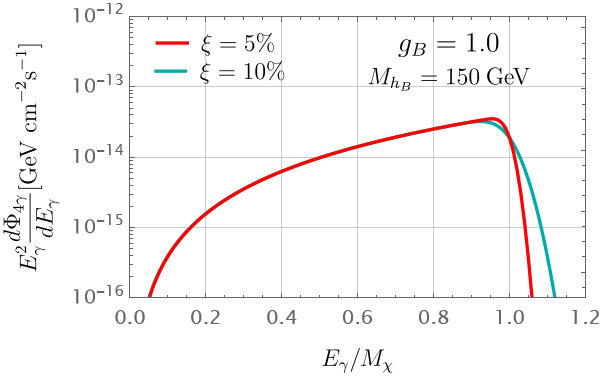}
         \caption{}
    \end{subfigure} 
        \caption{Photon flux from $\chi \chi \rightarrow h_B h_B \rightarrow 4 \gamma $ as a function of $E_\gamma /M_\chi$ and for different energy resolutions. In the left-panel we show the predictions in scenario I, while in the right-panel we show the predictions in scenario II.}
        \label{flux4gamma}
\end{figure}
In Fig.~\ref{flux4gamma} we show the differential flux for the channel producing four photons thanks to the fact that the new Higgs has a very large decay branching ratio into two photons. The values of the flux are small because the cross section for this annihilation channel is velocity suppressed.
Clearly, this process does not spoil the visibility of gamma lines in this theory.

\item $\chi \chi \to Z_B h_B \to q \bar{q} \gamma \gamma$:

As we have discussed in the previous section, in the most generic region of the parameter space, where one can achieve the correct relic density, the annihilation process $\chi \chi \to Z_B h_B$ is the dominant one. Since the decay branching ratio for $h_B \to \gamma \gamma$ is large when the new Higgs mass is below the $WW$ mass threshold, one can expect a large cross section for this mechanism producing photons that could spoil the visibility of gamma lines.
\begin{figure}
\begin{eqnarray*}    
\begin{gathered}
\begin{tikzpicture}[line width=1.5 pt,node distance=1 cm and 1.5 cm]
\coordinate[label = left: $\chi$] (i1);
\coordinate[below right = 2cm of i1](v1);
\coordinate[below left = 2cm of v1, label= left:$\chi$](i2);
\coordinate[above right = 2cm of v1] (f1);
\coordinate[above left = 0.1 cm of f1, label=left: $Z_B$];
\coordinate[below right =  2cm of v1] (f2);
\coordinate[below left = 0.6 cm of f2, label=$h_B$];
\coordinate[above right =  1.5cm of f2,label=right: $\gamma$] (f3);
\coordinate[below right =  1.5cm of f2,label=right: $\gamma$] (f4);
\coordinate[above right =  2cm of v1] (g2);
\coordinate[above right =  1.5cm of g2,label=right: $q$] (g3);
\coordinate[below right =  1.5cm of g2,label=right: $q$] (g4);
\draw[fermionnoarrow] (i1) -- (v1);
\draw[fermionnoarrow] (i2) -- (v1);
\draw[vector] (v1) -- (f1);
\draw[scalarnoarrow] (f2) -- (v1);
\draw[vector] (f3) -- (f2);
\draw[vector] (f2) -- (f4);
\draw[fermion] (g3) -- (g2);
\draw[fermion] (g2) -- (g4);
\draw[fill=gray] (v1) circle (.5cm);
\draw[fill=blue] (f1) circle (.1cm);
\draw[fill=red] (f2) circle (.1cm);
\end{tikzpicture}
\end{gathered} 
\end{eqnarray*}
\caption{$\chi \chi \rightarrow Z_B h_B \rightarrow q \bar{q} \gamma \gamma$.}
\label{hBZBdecays}
\end{figure}
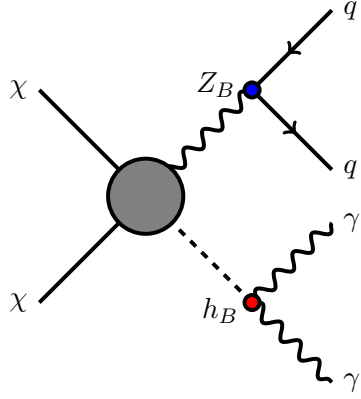
See Fig.~\ref{hBZBdecays} for the Feynman graph for this process.

In this case the energy of the photons is in the range:
\begin{equation}
E_\gamma^{min}=\frac{M_{h_B}}{2\gamma (1 + \beta)} \leq E_\gamma \leq  \frac{M_{h_B}}{2\gamma (1 - \beta)} =E_\gamma^{max},
\end{equation}
where 
\begin{eqnarray}
\gamma &=& \frac{E_{h_B}}{M_{h_B}}, \ \beta = \sqrt{1 - \frac{M_{h_B}^2}{E_{h_B}^2}},  \ \text{and} \
E_{h_B}= \frac{1}{ 4 M_\chi} \left( M_{h_B}^2 + 4 M_\chi^2 - M_{Z_B}^2\right).
\end{eqnarray}

\begin{figure}[h]
         \begin{subfigure}[b]{0.65\textwidth}
         \centering
         \includegraphics[width=\textwidth]{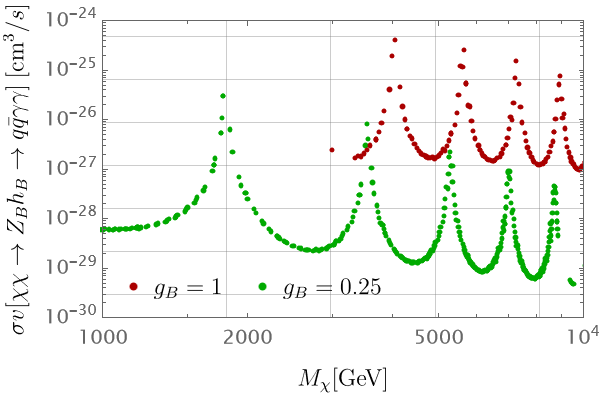}
         \caption{}
    \end{subfigure} 
        \caption{Annihilation cross section for the process $\chi \chi \to Z_B h_B \to q \bar{q} \gamma \gamma$. Here $M_{h_B}=150 $ GeV and $\sin{\theta_B}=0.01$.}
        \label{crossZBhB}
\end{figure}

In Fig.~\ref{crossZBhB} we show the numerical predictions for $\sigma v (\chi \chi \to Z_B h_B \to q \bar{q} \gamma \gamma)$ when $g_B=1$ (red dots) and $g_B=0.25$ (green dots). The Higgs mass is $M_{h_B}=150$ GeV, while the mixing angle between the Higgs is $\sin \theta_B=0.01$. As we expect the cross section for this process is large and could potential spoil the visibility of the gamma lines.

The flux for the photons from this annihilation channel is given by
\begin{equation}
\frac{d\Phi_{ \gamma }}{dE_\gamma}
=\frac{ 2 ( \sigma v_\text{rel} (\chi \chi \to Z_B h_B)) BR (h_B \to 2 \gamma)}{8\pi M_\chi^2}\frac{dN_{2\gamma}}{dE_\gamma} J_\text{ann},
\end{equation}
where
\begin{equation}
\frac{dN_{2 \gamma}}{dE_\gamma}=\int_0^\infty dE_0 \,  W_{2\gamma} \, G(E_\gamma, \xi /\omega, E_0),
\end{equation}
with
\begin{equation}
W_{2\gamma} = \frac{1}{(E^{max}_\gamma-E_\gamma^{min})}\Theta (E_\gamma - E_\gamma^{min}) \Theta (E_\gamma^{max} - E_\gamma).
\end{equation}
\begin{figure}[h]
         \centering
    \begin{subfigure}[b]{0.45\textwidth}
         \centering
         \includegraphics[width=\textwidth]{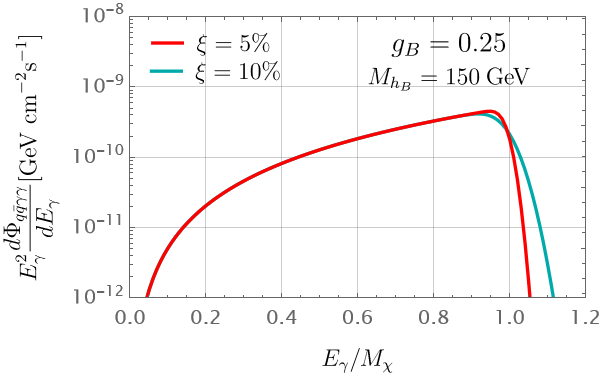}
         \caption{}
    \end{subfigure}  
         \begin{subfigure}[b]{0.45\textwidth}
         \centering
         \includegraphics[width=\textwidth]{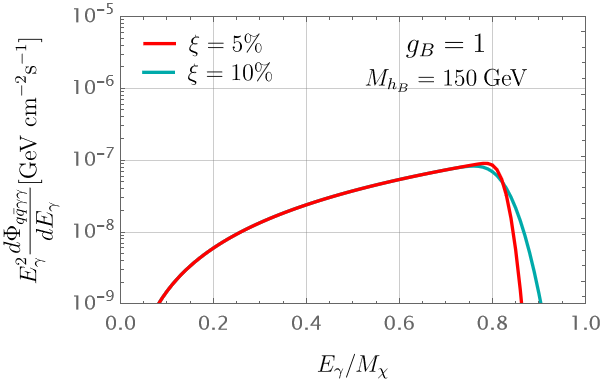}
         \caption{}
    \end{subfigure} 
        \caption{Differential flux from $\chi \chi \rightarrow \ZB h_B \rightarrow q\bar{q}\gamma \gamma $ as a function of $E_\gamma /M_\chi$ for different energy resolution. In the left-panel we show the predictions in scenario I, while in the right-panel we show the results in scenario II.}
        \label{fluxZBhB}
\end{figure}
In Fig.~\ref{fluxZBhB} we show the numerical for the differential flux for the photons produced in $\chi \chi \rightarrow \ZB h_B \rightarrow q\bar{q}\gamma \gamma$ in the two scenarios discussed above. As one can see, this process contributes to the continuum spectrum and can be close to the gamma line with energy equal to the dark matter mass. We will show that this process spoils the visibility of the gamma lines in some scenarios.

\item $\chi \chi \to q {\bar{q}} \gamma$:

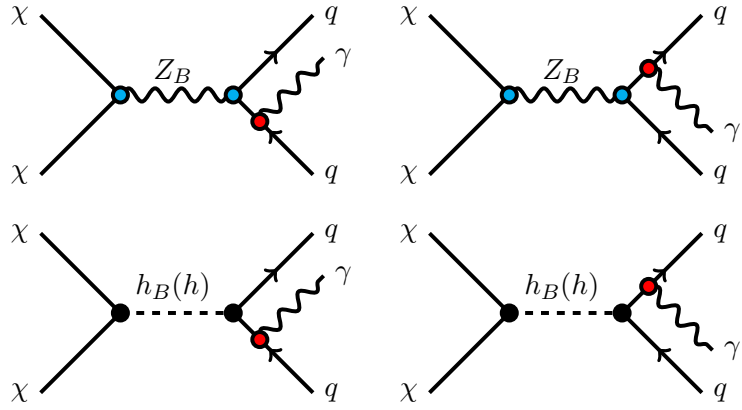
\begin{figure}
\begin{eqnarray*}
\begin{gathered}
\begin{tikzpicture}[line width=1.5 pt,node distance=1 cm and 1.5 cm]
\coordinate[label =left: $\chi$] (i1);
\coordinate[below right= 1.5cm of i1](v1);
\coordinate[right= 0.7cm of v1, label=$Z_B$](vaux);
\coordinate[below left= 1.5 cm of v1, label= left: $\chi$](i2);
\coordinate[right = 1.5 cm of v1](v2);
\coordinate[above right = 1.5 cm of v2, label=right: $q$] (f1);
\coordinate[below right =  1.5 cm of v2,label=right: $q$] (f2);
\coordinate[below right =  0.5 cm of v2](f3);
\coordinate[above right =  1.2 cm of f3, label=right: $\gamma$](f4);
\draw[fermionnoarrow] (i1) -- (v1);
\draw[fermionnoarrow] (i2) -- (v1);
\draw[vector] (v1) -- (v2);
\draw[fermion] (v2) -- (f1);
\draw[fermion] (f2) -- (v2);
\draw[vector] (f3) -- (f4);
\draw[fill=cyan] (v1) circle (.1cm);
\draw[fill=cyan] (v2) circle (.1cm);
\draw[fill=red] (f3) circle (.1cm);
\end{tikzpicture}
\end{gathered}
\quad
\begin{gathered}
\begin{tikzpicture}[line width=1.5 pt,node distance=1 cm and 1.5 cm]
\coordinate[label =left: $\chi$] (i1);
\coordinate[below right= 1.5cm of i1](v1);
\coordinate[right= 0.7cm of v1, label=$Z_B$](vaux);
\coordinate[below left= 1.5 cm of v1, label= left: $\chi$](i2);
\coordinate[right = 1.5 cm of v1](v2);
\coordinate[above right = 1.5 cm of v2, label=right: $q$] (f1);
\coordinate[below right =  1.5 cm of v2,label=right: $q$] (f2);
\coordinate[above right =  0.5 cm of v2](f3);
\coordinate[below right =  1.2 cm of f3, label=right: $\gamma$](f4);
\draw[fermionnoarrow] (i1) -- (v1);
\draw[fermionnoarrow] (i2) -- (v1);
\draw[vector] (v1) -- (v2);
\draw[fermion] (v2) -- (f1);
\draw[fermion] (f2) -- (v2);
\draw[vector] (f3) -- (f4);
\draw[fill=cyan] (v1) circle (.1cm);
\draw[fill=cyan] (v2) circle (.1cm);
\draw[fill=red] (f3) circle (.1cm);
\end{tikzpicture}
\end{gathered}\\
\begin{gathered}
\begin{tikzpicture}[line width=1.5 pt,node distance=1 cm and 1.5 cm]
\coordinate[label =left: $\chi$] (i1);
\coordinate[below right= 1.5cm of i1](v1);
\coordinate[right= 0.7cm of v1, label=$h_B(h)$](vaux);
\coordinate[below left= 1.5 cm of v1, label= left: $\chi$](i2);
\coordinate[right = 1.5 cm of v1](v2);
\coordinate[above right = 1.5 cm of v2, label=right: $q$] (f1);
\coordinate[below right =  1.5 cm of v2,label=right: $q$] (f2);
\coordinate[below right =  0.5 cm of v2](f3);
\coordinate[above right =  1.2 cm of f3, label=right: $\gamma$](f4);
\draw[fermionnoarrow] (i1) -- (v1);
\draw[fermionnoarrow] (i2) -- (v1);
\draw[scalarnoarrow] (v1) -- (v2);
\draw[fermion] (v2) -- (f1);
\draw[fermion] (f2) -- (v2);
\draw[vector] (f3) -- (f4);
\draw[fill=black] (v1) circle (.1cm);
\draw[fill=black] (v2) circle (.1cm);
\draw[fill=red] (f3) circle (.1cm);
\end{tikzpicture}
\end{gathered}
\quad
\begin{gathered}
\begin{tikzpicture}[line width=1.5 pt,node distance=1 cm and 1.5 cm]
\coordinate[label =left: $\chi$] (i1);
\coordinate[below right= 1.5cm of i1](v1);
\coordinate[right= 0.7cm of v1, label=$h_B(h)$](vaux);
\coordinate[below left= 1.5 cm of v1, label= left: $\chi$](i2);
\coordinate[right = 1.5 cm of v1](v2);
\coordinate[above right = 1.5 cm of v2, label=right: $q$] (f1);
\coordinate[below right =  1.5 cm of v2,label=right: $q$] (f2);
\coordinate[above right =  0.5 cm of v2](f3);
\coordinate[below right =  1.2 cm of f3, label=right: $\gamma$](f4);
\draw[fermionnoarrow] (i1) -- (v1);
\draw[fermionnoarrow] (i2) -- (v1);
\draw[scalarnoarrow] (v1) -- (v2);
\draw[fermion] (v2) -- (f1);
\draw[fermion] (f2) -- (v2);
\draw[vector] (f3) -- (f4);
\draw[fill=black] (v1) circle (.1cm);
\draw[fill=black] (v2) circle (.1cm);
\draw[fill=red] (f3) circle (.1cm);
\end{tikzpicture}
\end{gathered}
\end{eqnarray*}
\caption{Final state radiation processes mediated by the new gauge and Higgs bosons.}
\label{FSR}
\end{figure}
\begin{figure} [h]  
        \centering       
        \includegraphics[width=0.75\textwidth]{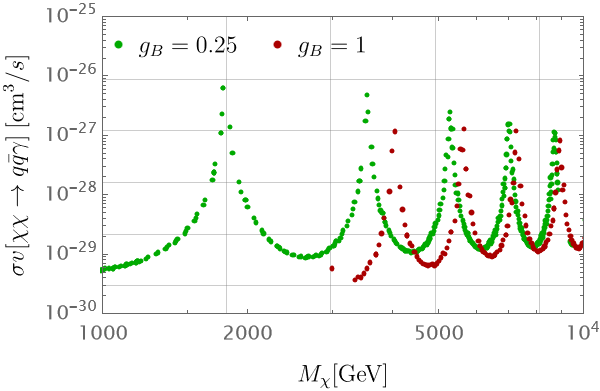}
         \caption{Fig shows $\sigma v(\chi \chi \rightarrow q \bar{q}\gamma) $ as a function of $M_{\chi}$ for different values of $g_B$. Here we used the corresponding values of $M_{Z_B}$ that satisfy the relic density constraints and $M_{h_B}$=150 GeV. }
         \label{crosssectionFSR}
\end{figure}
The final state radiation processes such as $\chi \ \chi \to q \ \bar{q} \ \gamma$ may spoil the visibility of the gamma lines. 
See the Feynman graphs for the final state radiation processes mediated by the new gauge and Higgs bosons in Fig.~\ref{FSR}.
The maximal energy of the photon in the final state radiation processes is given by
\begin{equation}
E^\gamma_{\rm{max}} = M_{\chi} \left(1 - \frac{M_{\rm{SM}}^2}{ M_{\chi}^2} \right),
\label{Eq2}
\end{equation}
where $\rm{M}_{\rm{SM}}$ is the mass of a Standard Model electric charged field. In this model, this process occurs at tree level while the gamma lines discussed above are quantum mechanical processes. The relevance of the final state radiation with respect to the gamma lines will be crucial to determine whether they can be observed or not. 
The amplitude squared of the processes contributing to the final state radiation (FSR) can be written as an expansion on the velocity~\cite{FileviezPerez:2019rcj}, 
\begin{equation}
|{\cal M}|^2_\text{FSR} = \frac{M_q^2}{M^2_{Z_B}} \frac{A}{4} +  v^2 \frac{B}{4} + {\cal O}(v^4),
\label{EqAp}
\end{equation}
where the leading order term of the expansion is suppressed by the ratio $(M_q/M_{Z_B})^2$. The coefficients in the above expression are given by
\begin{eqnarray}
&&A=- 12 \pi \, \alpha \, g_B^4  Q_q^2   (M_{Z_B}^2-4M_\chi^2)^2 \frac{\left(2(E_q-M_\chi)(E_q + E_\gamma -M_\chi)-3M_q^2 \right) }{M_{Z_B}^2(E_q-M_\chi)^2 ((4M_\chi^2-M_{Z_B}^2)^2+\Gamma_{Z_B}^2M_{Z_B}^2)},  \nonumber \\
&&B= 12 \pi \, \alpha \,  g_B^4 M_\chi^2 Q_q^2  \times \nonumber \\
&&  \frac{ \left(2E_q M_\chi (E_\gamma^2-3 E_\gamma M_\chi + 2 M_\chi^2)-2 E_q^4 - 2 E_q^3 ( E_\gamma - 2 M_\chi) -E_q^2(E_\gamma^2-6E_\gamma M_\chi + 6M_\chi^2)-2M_\chi^2(E_\gamma - M_\chi)^2\right)}
{E_q^2 (E_q - M_\chi) (E_q+E_\gamma-M_\chi)((4M_\chi^2-M_{Z_B}^2)^2+\Gamma_{Z_B}^2M_{Z_B}^2)}.\nonumber \end{eqnarray}
The DM annihilation to $\gamma \bar{q}q$ is described by a three-body phase space and the cross section is given by
\begin{equation}
\frac{d^2 \left( \sigma v_\text{rel} (\chi \chi \to q  \bar{q} \gamma)\right)}{dE_\gamma dE_f}=\frac{1}{32\pi^3 s}|{\cal M}|^2_\text{FSR}.
\end{equation}
In order to compute the annihilation cross section for final state radiation one needs to integrate the above equation with respect to $E_q$, which kinematic range is determined by the condition $\cos \theta_{q\gamma}^2 \leq 1 $. According to this, the integration limits are given by 
\begin{equation}
E_q^\pm = M_\chi - \frac{E_\gamma}{2} \pm \frac{E_\gamma}{2}\sqrt{1+\frac{M_q^2}{(E_\gamma - M_\chi)M_\chi}}.
\end{equation}
\begin{figure}[h]
         \centering
    \begin{subfigure}[b]{0.55\textwidth}
         \centering
         \includegraphics[width=\textwidth]{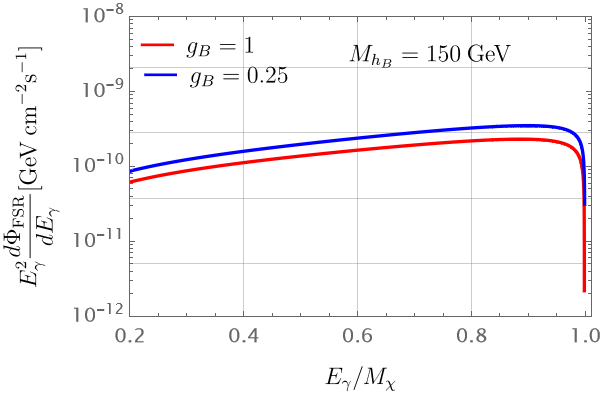}
    \end{subfigure}  
        \caption{Photon spectrum for $\chi \chi \rightarrow q\bar{q} \gamma $ as a function of $E_\gamma /M_\chi$ for different values of $g_B$. We show the results in scenario I when $g_B =0.25$ and scenario II when $g_B =1$.}
        \label{fluxFSR}
\end{figure}
Here we are mainly interested in the value of the FSR cross section close to the gamma lines to understand if the FSR spoils the visibility of the gamma lines. 

In Fig.~\ref{crosssectionFSR} we show the numerical values for $\sigma v(\chi \chi \rightarrow q \bar{q}\gamma) $ as a function of $M_{\chi}$ and for different values of $g_B$. Here we used the corresponding values of $M_{Z_B}$ that satisfy the relic density constraints and $M_{h_B}$=150 GeV. The cross section for these processes can be large in a large fraction of the part of the parameter space and can spoil the visibility of the gamma lines.

The flux for photons produced by the final state processes is given by
\begin{equation}
\frac{d\Phi_{FSR}}{dE_\gamma}= \frac{1}{8 \pi M_\chi^2} \frac{ d ( \sigma v_\text{rel} 
(\chi \chi \to q  \bar{q} \gamma))}{dE_\gamma}J_\text{ann},
\end{equation}
In Fig.~\ref{fluxFSR} we show the results for the different flux of the photons produced by the mechanisms in Fig.~\ref{FSR} in the scenarios I and II discussed above. The red (blue) line shows the predictions when $g_B=1$($g_B=0.25$). As one can appreciate, the flux can be large close to the energy of the gamma line with $E_\gamma=M_\chi$ and one could spoil the visibility of the gamma lines.
\end{itemize}
\begin{figure}[h]
         \centering
    \begin{subfigure}[b]{0.75\textwidth}
         \centering
         \includegraphics[width=\textwidth]{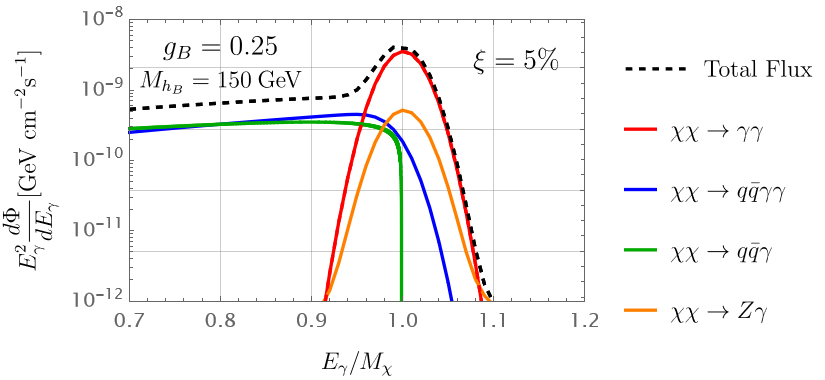}
         \caption{}
    \end{subfigure}  
         \begin{subfigure}[b]{0.75\textwidth}
         \centering
         \includegraphics[width=\textwidth]{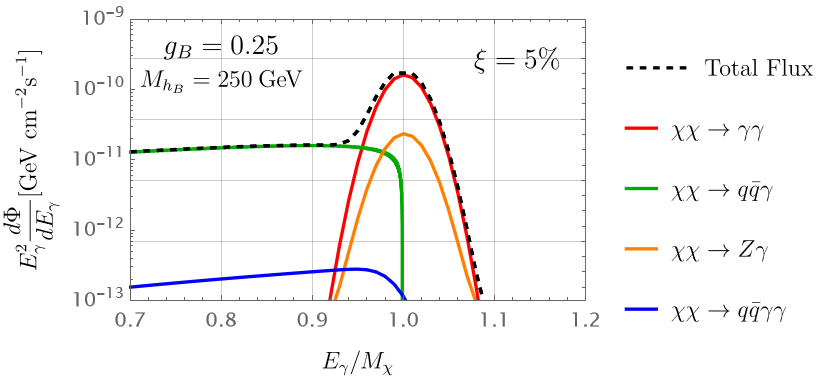}
         \caption{}
    \end{subfigure} 
        \caption{Photon flux as a function of $E_\gamma /M_\chi$ when the energy resolution is $5\%$ and $g_B =0.25$. In the upper-panel the energy spectrum is shown when  $M_{h_B}=150$ GeV, while in the lower-panel one has the spectrum when $M_{h_B}=250$ GeV. The rest of the parameters are defined in Scenario I.}
        \label{TOTESP1}
\end{figure}
Using the previous results we can discuss the visibility of the gamma lines comparing the ratio between the fluxes for gamma lines and continuum coming from final state radiation and the other processes. In Fig.~\ref{TOTESP1} we show the total flux in two cases: in the upper-panel $M_{h_B}=150$ GeV, while in the lower-panel $M_{h_B}=250$ GeV. In Fig.~\ref{TOTESP1} (a) we show the contribution to the total flux from the different processes discussed above. The red line shows the predictions from the gamma line, $\chi \chi \to \gamma \gamma$, the blue line corresponds to the predictions from $\chi \chi \to Z_B h_B \to q \bar{q} \gamma \gamma$, in green one has the predictions from the final state radiation processes, the orange shows the contribution from $\chi \chi \to Z \gamma$, and the total flux is shown by the black dashed line. As one can appreciate, the visibility of the gamma line is compromised due to the large contribution from $\chi \chi \to Z_B h_B \to q \bar{q} \gamma \gamma$ since the new Higgs with mass $150$ GeV has a large decay branching ratio into two photons. Now, in Fig.~\ref{TOTESP1} (b) the new Higgs mass is larger, $M_{h_B}=250$ GeV, and its decay branching ratio into two photons is very small. Therefore, the contribution to the photon total flux from the final state radiation is much larger than the contribution from $\chi \chi \to Z_B h_B \to q \bar{q} \gamma \gamma$. However, these contributions are much smaller than the contribution from the gamma line.
Therefore, in this case one can see very well the transition from the continuum to the gamma line.
\begin{figure}[h]
         \centering
    \begin{subfigure}[b]{0.75\textwidth}
         \centering
         \includegraphics[width=\textwidth]{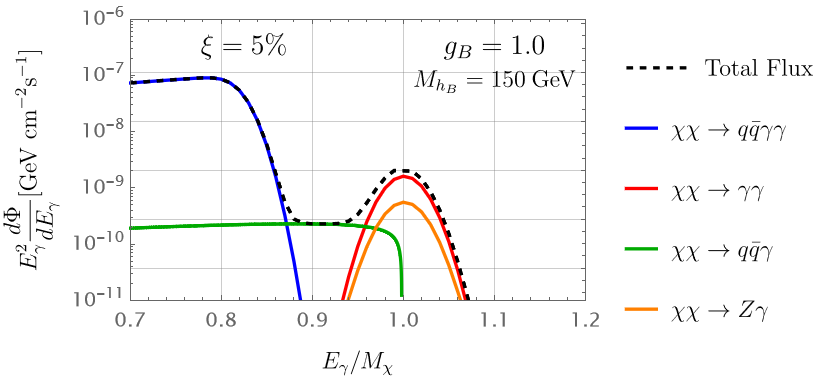}
         \caption{}
    \end{subfigure}  
         \begin{subfigure}[b]{0.75\textwidth}
         \centering
         \includegraphics[width=\textwidth]{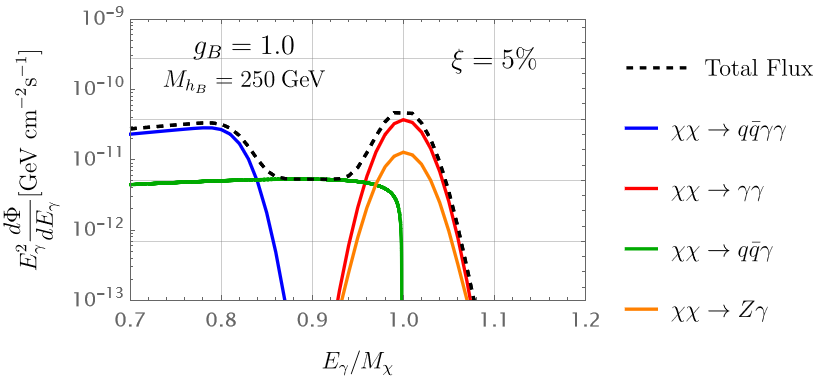}
         \caption{}
    \end{subfigure} 
        \caption{  Total Energy Spectrum  as a function of $E_\gamma /M_\chi$ for $5\%$ energy resolution when $g_B =1$. Fig (a) shows the energy spectrum when  $M_{h_B}=150$ GeV. Fig (b) shows the spectrum when $M_{h_B}=250$ GeV. The other parameters are defined as in Scenario II.}
        \label{TOTESP2}
\end{figure}

In Fig.~\ref{TOTESP2} we show the photon flux taking into all contributions when $g_B=1$ and the other parameters as in scenario II. In the upper-panel we have the results for $M_{h_B}=150$ GeV, while in the lower-panel one has $M_{h_B}=250$ GeV. In both cases 
the largest contributions are from the gamma-line and $\chi \chi \to Z_B h_B \to q \bar{q} \gamma \gamma$. When the Higgs mass is $M_{h_B}=150$ GeV (upper-panel), the $h_B$ decay branching ratio to photons is large and then the contribution to the photon flux from $\chi \chi \to Z_B h_B \to q \bar{q} \gamma \gamma$ is the largest one. When the Higgs mass is above the $WW$ mass threshold the branching ratio to photons is small, and the main contribution to the flux is from the gamma-line. Notice that even when the contribution from $\chi \chi \to Z_B h_B \to q \bar{q} \gamma \gamma$ is larger than the flux associated to the gamma-line one can see the transition from the continuum to the gamma-line. These features of the gamma spectrum are very interesting and unique for this dark matter theory. 
\section{SUMMARY}
\label{summary}
We discussed a class of theories where a dark matter candidate is predicted from gauge anomaly cancellation. The simplest theory based on local baryon number is investigated in detail. This theory predicts a Majorana dark matter candidate with fractional baryon number which stability is a natural consequence from symmetry breaking.
We discussed the dark matter candidates and the correlation between the relic density constraints and the predictions for direct detection. We studied the new Higgs decays showing that its branching ratio into two photons can be large. 
The allowed parameter space has been discussed to understand the testability of this theory for dark matter in different experiments.

We have discussed in great detail the predictions for gamma lines and the processes predicting a continuum spectrum for the photons produced from dark matter annihilation. In this theory we have two type of gamma lines: $\chi \chi \to Z_B^* \to \gamma \gamma$ and $\chi \chi \to Z_B^* \to Z \gamma$. The needed $Z_B \gamma \gamma$ and $Z_B Z \gamma$ effective couplings are predicted in this anomaly-free theory for dark matter where inside the loop one has the contributions of the new fermions needed for anomaly cancellation.
We have discussed the processes, $\chi \chi \to h_B h_B \to \gamma \gamma \gamma \gamma$ and $\chi \chi \to Z_B h_B \to q \bar{q} \gamma \gamma$. These processes are important because the decay branching ratio of the new Higgs into two photons is very large when the Higgs mass is below the $WW$ mass threshold and the mixing angle is small as required by the direct detection bounds. We have investigated the final state radiation contribution, $\chi \chi \to Z_B \to q \bar{q} \gamma$, to the photon flux. We have shown that the predictions for the gamma line produced in $\chi \chi \to Z \gamma$ can be tested in the near future at gamma-ray telescope such as CTA. We have discussed the predictions for the total photon flux in several scenarios, showing that the unique features of the gamma-ray spectrum can be used to test this theory for dark matter. 

{\textit{Acknowledgements}:} P.F.P. thanks the SIMONS Foundation for financial support and the Galileo Galilei Institute for Theoretical Physics in Florence, Italy, for hospitality. We thank the referee for very useful suggestions. This work made use of the High Performance Computing Resource in the Core Facility for Advanced Research Computing at Case Western Reserve University.
\newpage
\appendix

\section{DARK MATTER ANNIHILATION CHANNELS}
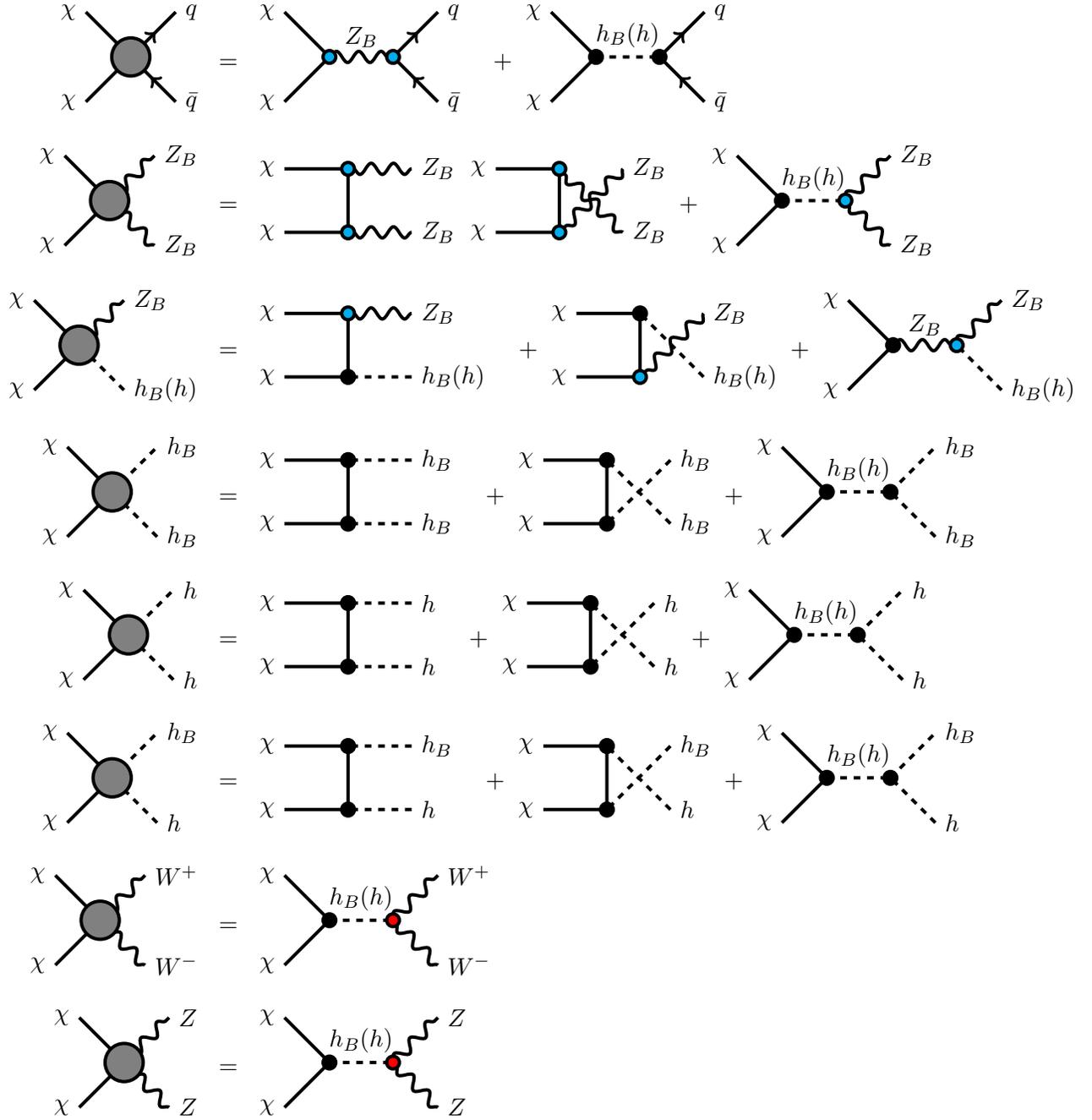
\begin{figure}[h]
\begin{eqnarray*}
\begin{gathered}
\begin{tikzpicture}[line width=1.5 pt,node distance=1 cm and 1.5 cm]
\coordinate[label = left: $\chi$] (i1);
\coordinate[below right = 1cm of i1](v1);
\coordinate[below left = 1cm of v1, label= left:$\chi$](i2);
\coordinate[above right = 1cm of v1, label=right: $q$] (f1);
\coordinate[below right =  1cm of v1,label=right: $\bar{q}$] (f2);
\draw[fermionnoarrow] (i1) -- (v1);
\draw[fermionnoarrow] (i2) -- (v1);
\draw[fermion] (v1) -- (f1);
\draw[fermion] (f2) -- (v1);
\draw[fill=gray] (v1) circle (.3cm);
\end{tikzpicture}
\end{gathered} 
&=&
\begin{gathered}
\begin{tikzpicture}[line width=1.5 pt,node distance=1 cm and 1.5 cm]
\coordinate[label =left: $\chi$] (i1);
\coordinate[below right= 1cm of i1](v1);
\coordinate[ right= 0.5cm of v1,label= above:$Z_B$](vaux);
\coordinate[below left= 1cm of v1, label= left: $\chi$](i2);
\coordinate[right = 1 cm of v1](v2);
\coordinate[above right = 1 cm of v2, label=right: $q$] (f1);
\coordinate[below right =  1 cm of v2,label=right: $\bar{q}$] (f2);
\draw[fermionnoarrow] (i1) -- (v1);
\draw[fermionnoarrow] (i2) -- (v1);
\draw[vector] (v1) -- (v2);
\draw[fermion] (v2) -- (f1);
\draw[fermion] (f2) -- (v2);
\draw[fill=cyan] (v1) circle (.1cm);
\draw[fill=cyan] (v2) circle (.1cm);
\end{tikzpicture}
\end{gathered}
\quad   \! +
\begin{gathered}
\begin{tikzpicture}[line width=1.5 pt,node distance=1 cm and 1.5 cm]
\coordinate[label =left: $\chi$] (i1);
\coordinate[below right= 1cm of i1](v1);
\coordinate[ right= 0.5cm of v1,label= above:$h_B(h)$](vaux);
\coordinate[below left= 1cm of v1, label= left: $\chi$](i2);
\coordinate[right = 1 cm of v1](v2);
\coordinate[above right = 1 cm of v2, label=right: $q$] (f1);
\coordinate[below right =  1 cm of v2,label=right: $\bar{q}$] (f2);
\draw[fermionnoarrow] (i1) -- (v1);
\draw[fermionnoarrow] (i2) -- (v1);
\draw[scalarnoarrow] (v1) -- (v2);
\draw[fermion] (v2) -- (f1);
\draw[fermion] (f2) -- (v2);
\draw[fill=black] (v1) circle (.1cm);
\draw[fill=black] (v2) circle (.1cm);
\end{tikzpicture}
\end{gathered}\\
\begin{gathered}
\begin{tikzpicture}[line width=1.5 pt,node distance=1 cm and 1.5 cm]
\coordinate[label = left: $\chi$] (i1);
\coordinate[below right = 1cm of i1](v1);
\coordinate[below left = 1cm of v1, label= left:$\chi$](i2);
\coordinate[above right = 1cm of v1, label=right: $Z_B$] (f1);
\coordinate[below right =  1cm of v1,label=right: $Z_B$] (f2);
\draw[fermionnoarrow] (i1) -- (v1);
\draw[fermionnoarrow] (i2) -- (v1);
\draw[vector] (v1) -- (f1);
\draw[vector] (f2) -- (v1);
\draw[fill=gray] (v1) circle (.3cm);
\end{tikzpicture}
\end{gathered} 
&=&
\begin{gathered}
\begin{tikzpicture}[line width=1.5 pt,node distance=1 cm and 1.5 cm]
\coordinate[label =left: $\chi$] (i1);
\coordinate[right= 1cm of i1](v1);
\coordinate[below= 0.5cm of v1](vaux);
\coordinate[right = 1cm of v1, label= right:$Z_B$](f1);
\coordinate[below = 1 cm of v1](v2);
\coordinate[left = 1 cm of v2, label=left: $\chi$] (i2);
\coordinate[right =  1 cm of v2,label=right: $Z_B$] (f2);
\draw[fermionnoarrow] (i1) -- (v1);
\draw[fermionnoarrow] (v1) -- (v2);
\draw[fermionnoarrow] (i2) -- (v2);
\draw[vector] (f2) -- (v2);
\draw[vector] (f1) -- (v1);
\draw[fill=cyan] (v1) circle (.1cm);
\draw[fill=cyan] (v2) circle (.1cm);
\end{tikzpicture}
\end{gathered}
\begin{gathered}
\begin{tikzpicture}[line width=1.5 pt,node distance=1 cm and 1.5 cm]
\coordinate[label =left: $\chi$] (i1);
\coordinate[right= 1cm of i1](v1);
\coordinate[below= 0.5cm of v1,label](vaux);
\coordinate[right = 1cm of v1, label= right:$Z_B$](f1);
\coordinate[below = 1 cm of v1](v2);
\coordinate[left = 1 cm of v2, label=left: $\chi$] (i2);
\coordinate[right =  1 cm of v2,label=right: $Z_B$] (f2);
\draw[fermionnoarrow] (i1) -- (v1);
\draw[fermionnoarrow] (v1) -- (v2);
\draw[fermionnoarrow] (i2) -- (v2);
\draw[vector] (f2) -- (v1);
\draw[vector] (f1) -- (v2);
\draw[fill=cyan] (v1) circle (.1cm);
\draw[fill=cyan] (v2) circle (.1cm);
\end{tikzpicture}
\end{gathered}
+
\begin{gathered}
\begin{tikzpicture}[line width=1.5 pt,node distance=1 cm and 1.5 cm]
\coordinate[label =left: $\chi$] (i1);
\coordinate[below right= 1cm of i1](v1);
\coordinate[ right= 0.5cm of v1,label= above:$h_B (h)$](vaux);
\coordinate[below left= 1cm of v1, label= left: $\chi$](i2);
\coordinate[right = 1 cm of v1](v2);
\coordinate[above right = 1 cm of v2, label=right: $Z_B$] (f1);
\coordinate[below right =  1 cm of v2,label=right: $Z_B$] (f2);
\draw[fermionnoarrow] (i1) -- (v1);
\draw[fermionnoarrow] (i2) -- (v1);
\draw[scalarnoarrow] (v1) -- (v2);
\draw[vector] (v2) -- (f1);
\draw[vector] (f2) -- (v2);
\draw[fill=black] (v1) circle (.1cm);
\draw[fill=cyan] (v2) circle (.1cm);
\end{tikzpicture}
\end{gathered} \\
\begin{gathered}
\begin{tikzpicture}[line width=1.5 pt,node distance=1 cm and 1.5 cm]
\coordinate[label = left: $\chi$] (i1);
\coordinate[below right = 1cm of i1](v1);
\coordinate[below left = 1cm of v1, label= left:$\chi$](i2);
\coordinate[above right = 1cm of v1, label=right: $Z_B$] (f1);
\coordinate[below right =  1cm of v1,label=right: $h_B(h)$] (f2);
\draw[fermionnoarrow] (i1) -- (v1);
\draw[fermionnoarrow] (i2) -- (v1);
\draw[vector] (v1) -- (f1);
\draw[scalarnoarrow] (f2) -- (v1);
\draw[fill=gray] (v1) circle (.3cm);
\end{tikzpicture}
\end{gathered} 
&=&
\begin{gathered}
\begin{tikzpicture}[line width=1.5 pt,node distance=1 cm and 1.5 cm]
\coordinate[label =left: $\chi$] (i1);
\coordinate[right= 1cm of i1](v1);
\coordinate[below= 0.5cm of v1](vaux);
\coordinate[right = 1cm of v1, label= right:$Z_B$](f1);
\coordinate[below = 1 cm of v1](v2);
\coordinate[left = 1 cm of v2, label=left: $\chi$] (i2);
\coordinate[right =  1 cm of v2,label=right: $h_B(h)$] (f2);
\draw[fermionnoarrow] (i1) -- (v1);
\draw[fermionnoarrow] (v1) -- (v2);
\draw[fermionnoarrow] (i2) -- (v2);
\draw[scalarnoarrow] (f2) -- (v2);
\draw[vector] (f1) -- (v1);
\draw[fill=cyan] (v1) circle (.1cm);
\draw[fill=black] (v2) circle (.1cm);
\end{tikzpicture}
\end{gathered}
\quad   \! +
\begin{gathered}
\begin{tikzpicture}[line width=1.5 pt,node distance=1 cm and 1.5 cm]
\coordinate[label =left: $\chi$] (i1);
\coordinate[right= 1cm of i1](v1);
\coordinate[below= 0.5cm of v1,label](vaux);
\coordinate[right = 1cm of v1, label= right:$Z_B$](f1);
\coordinate[below = 1 cm of v1](v2);
\coordinate[left = 1 cm of v2, label=left: $\chi$] (i2);
\coordinate[right =  1 cm of v2,label=right: $h_B(h)$] (f2);
\draw[fermionnoarrow] (i1) -- (v1);
\draw[fermionnoarrow] (v1) -- (v2);
\draw[fermionnoarrow] (i2) -- (v2);
\draw[scalarnoarrow] (f2) -- (v1);
\draw[vector] (f1) -- (v2);
\draw[fill=black] (v1) circle (.1cm);
\draw[fill=cyan] (v2) circle (.1cm);
\end{tikzpicture}
\end{gathered}
+
\begin{gathered}
\begin{tikzpicture}[line width=1.5 pt,node distance=1 cm and 1.5 cm]
\coordinate[label =left: $\chi$] (i1);
\coordinate[below right= 1cm of i1](v1);
\coordinate[ right= 0.5cm of v1,label= above:$Z_B$](vaux);
\coordinate[below left= 1cm of v1, label= left: $\chi$](i2);
\coordinate[right = 1 cm of v1](v2);
\coordinate[above right = 1 cm of v2, label=right: $Z_B$] (f1);
\coordinate[below right =  1 cm of v2,label=right: $h_B(h)$] (f2);
\draw[fermionnoarrow] (i1) -- (v1);
\draw[fermionnoarrow] (i2) -- (v1);
\draw[vector] (v1) -- (v2);
\draw[vector] (v2) -- (f1);
\draw[scalarnoarrow] (f2) -- (v2);
\draw[fill=black] (v1) circle (.1cm);
\draw[fill=cyan] (v2) circle (.1cm);
\end{tikzpicture}
\end{gathered} \\
\begin{gathered}
\begin{tikzpicture}[line width=1.5 pt,node distance=1 cm and 1.5 cm]
\coordinate[label = left: $\chi$] (i1);
\coordinate[below right = 1cm of i1](v1);
\coordinate[below left = 1cm of v1, label= left:$\chi$](i2);
\coordinate[above right = 1cm of v1, label=right: $h_B$] (f1);
\coordinate[below right =  1cm of v1,label=right: $h_B$] (f2);
\draw[fermionnoarrow] (i1) -- (v1);
\draw[fermionnoarrow] (i2) -- (v1);
\draw[scalarnoarrow] (v1) -- (f1);
\draw[scalarnoarrow] (f2) -- (v1);
\draw[fill=gray] (v1) circle (.3cm);
\end{tikzpicture}
\end{gathered} 
&=&
\begin{gathered}
\begin{tikzpicture}[line width=1.5 pt,node distance=1 cm and 1.5 cm]
\coordinate[label =left: $\chi$] (i1);
\coordinate[right= 1cm of i1](v1);
\coordinate[below= 0.5cm of v1](vaux);
\coordinate[right = 1cm of v1, label= right:$h_B$](f1);
\coordinate[below = 1 cm of v1](v2);
\coordinate[left = 1 cm of v2, label=left: $\chi$] (i2);
\coordinate[right =  1 cm of v2,label=right: $h_B$] (f2);
\draw[fermionnoarrow] (i1) -- (v1);
\draw[fermionnoarrow] (v1) -- (v2);
\draw[fermionnoarrow] (i2) -- (v2);
\draw[scalarnoarrow] (f2) -- (v2);
\draw[scalarnoarrow] (f1) -- (v1);
\draw[fill=black] (v1) circle (.1cm);
\draw[fill=black] (v2) circle (.1cm);
\end{tikzpicture}
\end{gathered}
\quad   \! +
\begin{gathered}
\begin{tikzpicture}[line width=1.5 pt,node distance=1 cm and 1.5 cm]
\coordinate[label =left: $\chi$] (i1);
\coordinate[right= 1cm of i1](v1);
\coordinate[below= 0.5cm of v1,label](vaux);
\coordinate[right = 1cm of v1, label= right:$h_B$](f1);
\coordinate[below = 1 cm of v1](v2);
\coordinate[left = 1 cm of v2, label=left: $\chi$] (i2);
\coordinate[right =  1 cm of v2,label=right: $h_B$] (f2);
\draw[fermionnoarrow] (i1) -- (v1);
\draw[fermionnoarrow] (v1) -- (v2);
\draw[fermionnoarrow] (i2) -- (v2);
\draw[scalarnoarrow] (f2) -- (v1);
\draw[scalarnoarrow] (f1) -- (v2);
\draw[fill=black] (v1) circle (.1cm);
\draw[fill=black] (v2) circle (.1cm);
\end{tikzpicture}
\end{gathered}
+
\begin{gathered}
\begin{tikzpicture}[line width=1.5 pt,node distance=1 cm and 1.5 cm]
\coordinate[label =left: $\chi$] (i1);
\coordinate[below right= 1cm of i1](v1);
\coordinate[ right= 0.5cm of v1,label= above:$h_B(h)$](vaux);
\coordinate[below left= 1cm of v1, label= left: $\chi$](i2);
\coordinate[right = 1 cm of v1](v2);
\coordinate[above right = 1 cm of v2, label=right: $h_B$] (f1);
\coordinate[below right =  1 cm of v2,label=right: $h_B$] (f2);
\draw[fermionnoarrow] (i1) -- (v1);
\draw[fermionnoarrow] (i2) -- (v1);
\draw[scalarnoarrow] (v1) -- (v2);
\draw[scalarnoarrow] (v2) -- (f1);
\draw[scalarnoarrow] (f2) -- (v2);
\draw[fill=black] (v1) circle (.1cm);
\draw[fill=black] (v2) circle (.1cm);
\end{tikzpicture}
\end{gathered}\\
\begin{gathered}
\begin{tikzpicture}[line width=1.5 pt,node distance=1 cm and 1.5 cm]
\coordinate[label = left: $\chi$] (i1);
\coordinate[below right = 1cm of i1](v1);
\coordinate[below left = 1cm of v1, label= left:$\chi$](i2);
\coordinate[above right = 1cm of v1, label=right: $h$] (f1);
\coordinate[below right =  1cm of v1,label=right: $h$] (f2);
\draw[fermionnoarrow] (i1) -- (v1);
\draw[fermionnoarrow] (i2) -- (v1);
\draw[scalarnoarrow] (v1) -- (f1);
\draw[scalarnoarrow] (f2) -- (v1);
\draw[fill=gray] (v1) circle (.3cm);
\end{tikzpicture}
\end{gathered} 
&=&
\begin{gathered}
\begin{tikzpicture}[line width=1.5 pt,node distance=1 cm and 1.5 cm]
\coordinate[label =left: $\chi$] (i1);
\coordinate[right= 1cm of i1](v1);
\coordinate[below= 0.5cm of v1](vaux);
\coordinate[right = 1cm of v1, label= right:$h$](f1);
\coordinate[below = 1 cm of v1](v2);
\coordinate[left = 1 cm of v2, label=left: $\chi$] (i2);
\coordinate[right =  1 cm of v2,label=right: $h$] (f2);
\draw[fermionnoarrow] (i1) -- (v1);
\draw[fermionnoarrow] (v1) -- (v2);
\draw[fermionnoarrow] (i2) -- (v2);
\draw[scalarnoarrow] (f2) -- (v2);
\draw[scalarnoarrow] (f1) -- (v1);
\draw[fill=black] (v1) circle (.1cm);
\draw[fill=black] (v2) circle (.1cm);
\end{tikzpicture}
\end{gathered}
\quad   \! +
\begin{gathered}
\begin{tikzpicture}[line width=1.5 pt,node distance=1 cm and 1.5 cm]
\coordinate[label =left: $\chi$] (i1);
\coordinate[right= 1cm of i1](v1);
\coordinate[below= 0.5cm of v1,label](vaux);
\coordinate[right = 1cm of v1, label= right:$h$](f1);
\coordinate[below = 1 cm of v1](v2);
\coordinate[left = 1 cm of v2, label=left: $\chi$] (i2);
\coordinate[right =  1 cm of v2,label=right: $h$] (f2);
\draw[fermionnoarrow] (i1) -- (v1);
\draw[fermionnoarrow] (v1) -- (v2);
\draw[fermionnoarrow] (i2) -- (v2);
\draw[scalarnoarrow] (f2) -- (v1);
\draw[scalarnoarrow] (f1) -- (v2);
\draw[fill=black] (v1) circle (.1cm);
\draw[fill=black] (v2) circle (.1cm);
\end{tikzpicture}
\end{gathered}
+
\begin{gathered}
\begin{tikzpicture}[line width=1.5 pt,node distance=1 cm and 1.5 cm]
\coordinate[label =left: $\chi$] (i1);
\coordinate[below right= 1cm of i1](v1);
\coordinate[ right= 0.5cm of v1,label= above:$h_B(h)$](vaux);
\coordinate[below left= 1cm of v1, label= left: $\chi$](i2);
\coordinate[right = 1 cm of v1](v2);
\coordinate[above right = 1 cm of v2, label=right: $h$] (f1);
\coordinate[below right =  1 cm of v2,label=right: $h$] (f2);
\draw[fermionnoarrow] (i1) -- (v1);
\draw[fermionnoarrow] (i2) -- (v1);
\draw[scalarnoarrow] (v1) -- (v2);
\draw[scalarnoarrow] (v2) -- (f1);
\draw[scalarnoarrow] (f2) -- (v2);
\draw[fill=black] (v1) circle (.1cm);
\draw[fill=black] (v2) circle (.1cm);
\end{tikzpicture}
\end{gathered}\\
\begin{gathered}
\begin{tikzpicture}[line width=1.5 pt,node distance=1 cm and 1.5 cm]
\coordinate[label = left: $\chi$] (i1);
\coordinate[below right = 1cm of i1](v1);
\coordinate[below left = 1cm of v1, label= left:$\chi$](i2);
\coordinate[above right = 1cm of v1, label=right: $h_B$] (f1);
\coordinate[below right =  1cm of v1,label=right: $h$] (f2);
\draw[fermionnoarrow] (i1) -- (v1);
\draw[fermionnoarrow] (i2) -- (v1);
\draw[scalarnoarrow] (v1) -- (f1);
\draw[scalarnoarrow] (f2) -- (v1);
\draw[fill=gray] (v1) circle (.3cm);
\end{tikzpicture}
\end{gathered} 
&=&
\begin{gathered}
\begin{tikzpicture}[line width=1.5 pt,node distance=1 cm and 1.5 cm]
\coordinate[label =left: $\chi$] (i1);
\coordinate[right= 1cm of i1](v1);
\coordinate[below= 0.5cm of v1](vaux);
\coordinate[right = 1cm of v1, label= right:$h_B$](f1);
\coordinate[below = 1 cm of v1](v2);
\coordinate[left = 1 cm of v2, label=left: $\chi$] (i2);
\coordinate[right =  1 cm of v2,label=right: $h$] (f2);
\draw[fermionnoarrow] (i1) -- (v1);
\draw[fermionnoarrow] (v1) -- (v2);
\draw[fermionnoarrow] (i2) -- (v2);
\draw[scalarnoarrow] (f2) -- (v2);
\draw[scalarnoarrow] (f1) -- (v1);
\draw[fill=black] (v1) circle (.1cm);
\draw[fill=black] (v2) circle (.1cm);
\end{tikzpicture}
\end{gathered}
\quad   \! +
\begin{gathered}
\begin{tikzpicture}[line width=1.5 pt,node distance=1 cm and 1.5 cm]
\coordinate[label =left: $\chi$] (i1);
\coordinate[right= 1cm of i1](v1);
\coordinate[below= 0.5cm of v1,label](vaux);
\coordinate[right = 1cm of v1, label= right:$h_B$](f1);
\coordinate[below = 1 cm of v1](v2);
\coordinate[left = 1 cm of v2, label=left: $\chi$] (i2);
\coordinate[right =  1 cm of v2,label=right: $h$] (f2);
\draw[fermionnoarrow] (i1) -- (v1);
\draw[fermionnoarrow] (v1) -- (v2);
\draw[fermionnoarrow] (i2) -- (v2);
\draw[scalarnoarrow] (f2) -- (v1);
\draw[scalarnoarrow] (f1) -- (v2);
\draw[fill=black] (v1) circle (.1cm);
\draw[fill=black] (v2) circle (.1cm);
\end{tikzpicture}
\end{gathered}
+
\begin{gathered}
\begin{tikzpicture}[line width=1.5 pt,node distance=1 cm and 1.5 cm]
\coordinate[label =left: $\chi$] (i1);
\coordinate[below right= 1cm of i1](v1);
\coordinate[ right= 0.5cm of v1,label= above:$h_B(h)$](vaux);
\coordinate[below left= 1cm of v1, label= left: $\chi$](i2);
\coordinate[right = 1 cm of v1](v2);
\coordinate[above right = 1 cm of v2, label=right: $h_B$] (f1);
\coordinate[below right =  1 cm of v2,label=right: $h$] (f2);
\draw[fermionnoarrow] (i1) -- (v1);
\draw[fermionnoarrow] (i2) -- (v1);
\draw[scalarnoarrow] (v1) -- (v2);
\draw[scalarnoarrow] (v2) -- (f1);
\draw[scalarnoarrow] (f2) -- (v2);
\draw[fill=black] (v1) circle (.1cm);
\draw[fill=black] (v2) circle (.1cm);
\end{tikzpicture}
\end{gathered}\\
\begin{gathered}
\begin{tikzpicture}[line width=1.5 pt,node distance=1 cm and 1.5 cm]
\coordinate[label = left: $\chi$] (i1);
\coordinate[below right = 1cm of i1](v1);
\coordinate[below left = 1cm of v1, label= left:$\chi$](i2);
\coordinate[above right = 1cm of v1, label=right: $W^+$] (f1);
\coordinate[below right =  1cm of v1,label=right: $W^-$] (f2);
\draw[fermionnoarrow] (i1) -- (v1);
\draw[fermionnoarrow] (i2) -- (v1);
\draw[vector] (v1) -- (f1);
\draw[vector] (f2) -- (v1);
\draw[fill=gray] (v1) circle (.3cm);
\end{tikzpicture}
\end{gathered} 
&=&
\begin{gathered}
\begin{tikzpicture}[line width=1.5 pt,node distance=1 cm and 1.5 cm]
\coordinate[label =left: $\chi$] (i1);
\coordinate[below right= 1cm of i1](v1);
\coordinate[ right= 0.5cm of v1,label= above:$h_B (h)$](vaux);
\coordinate[below left= 1cm of v1, label= left: $\chi$](i2);
\coordinate[right = 1 cm of v1](v2);
\coordinate[above right = 1 cm of v2, label=right: $W^+$] (f1);
\coordinate[below right =  1 cm of v2,label=right: $W^-$] (f2);
\draw[fermionnoarrow] (i1) -- (v1);
\draw[fermionnoarrow] (i2) -- (v1);
\draw[scalarnoarrow] (v1) -- (v2);
\draw[vector] (v2) -- (f1);
\draw[vector] (f2) -- (v2);
\draw[fill=black] (v1) circle (.1cm);
\draw[fill=red] (v2) circle (.1cm);
\end{tikzpicture}
\end{gathered} \\
\begin{gathered}
\begin{tikzpicture}[line width=1.5 pt,node distance=1 cm and 1.5 cm]
\coordinate[label = left: $\chi$] (i1);
\coordinate[below right = 1cm of i1](v1);
\coordinate[below left = 1cm of v1, label= left:$\chi$](i2);
\coordinate[above right = 1cm of v1, label=right: $Z$] (f1);
\coordinate[below right =  1cm of v1,label=right: $Z$] (f2);
\draw[fermionnoarrow] (i1) -- (v1);
\draw[fermionnoarrow] (i2) -- (v1);
\draw[vector] (v1) -- (f1);
\draw[vector] (f2) -- (v1);
\draw[fill=gray] (v1) circle (.3cm);
\end{tikzpicture}
\end{gathered} 
&=&
\begin{gathered}
\begin{tikzpicture}[line width=1.5 pt,node distance=1 cm and 1.5 cm]
\coordinate[label =left: $\chi$] (i1);
\coordinate[below right= 1cm of i1](v1);
\coordinate[ right= 0.5cm of v1,label= above:$h_B (h)$](vaux);
\coordinate[below left= 1cm of v1, label= left: $\chi$](i2);
\coordinate[right = 1 cm of v1](v2);
\coordinate[above right = 1 cm of v2, label=right: $Z$] (f1);
\coordinate[below right =  1 cm of v2,label=right: $Z$] (f2);
\draw[fermionnoarrow] (i1) -- (v1);
\draw[fermionnoarrow] (i2) -- (v1);
\draw[scalarnoarrow] (v1) -- (v2);
\draw[vector] (v2) -- (f1);
\draw[vector] (f2) -- (v2);
\draw[fill=black] (v1) circle (.1cm);
\draw[fill=red] (v2) circle (.1cm);
\end{tikzpicture}
\end{gathered}
\end{eqnarray*}
\caption{Feynman graphs for the dark matter annihilation channels.}
\label{annihilation}
\end{figure}
\newpage
\section{FEYNMAN RULES}
\begin{align}
    q \bar{q} \ZB  &: \hspace{0.5cm}  i\frac{\gB}{3} \gamma^\mu, \\
    \chi \chi Z_B &: - i \frac{3}{4} g_B \gamma^\mu \gamma^5, \\
    \chi \chi h &: - i \frac{3}{2} \frac{g_B M_\chi}{M_{Z_B}} \sin \theta_B, \\
     \chi \chi h_B &: i \frac{3}{2} \frac{g_B M_\chi}{M_{Z_B}} \cos \theta_B, \\
     h_B Z_B Z_B &: - i {3} g_B M_{Z_B} \cos \theta_B \ g^{\mu \nu}, \\
     h Z_B Z_B &:  i 3 g_B M_{Z_B} \sin \theta_B \ g^{\mu \nu}, \\
     h Z Z & : \hspace{0.5 cm} 2i  \frac{M_{Z}^2}{v_0} \cos{\tB} \ g^{\mu \nu},\\
      \HB Z Z & : \hspace{0.5 cm}  2i \frac{M_{Z}^2}{v_0} \sin{\tB} \ g^{\mu \nu}, \\
      h W W & : \hspace{0.5 cm} 2i   \frac{M_{W}^2}{v_0} \cos{\tB} \ g^{\mu \nu},  \\
      \HB W W & : \hspace{0.5 cm}  2i   \frac{M_{W}^2}{v_0} \sin{\tB} \ g^{\mu \nu}, \\
      {\overline{\rho^-} \rho^- Z_B} &: {i \frac{3}{4} g_B \gamma^\mu \gamma^5},\\
      \overline{\Psi^-} \Psi^- Z_B  &: -i \frac{3}{4} g_B  \gamma^\mu \gamma^5, \\
      {\overline{\rho^-} \rho^- h_B} &: {i \frac{3}{2} \frac{g_B M_{\rho^-}}{M_{Z_B}}  \cos{\tB}},\\
      {\overline{\Psi^-} \Psi^- h_B} &: {i \frac{3}{2} \frac{g_B M_{\Psi^-}}{M_{Z_B}} \cos{\tB}}.
\end{align}
\newpage
\section{RELIC DENSITY RESULTS}
Since in the previous sections we used the scenario II where $g_B=1.0$, we show here the constraints from the relic density bounds for this scenario:
\begin{figure} [h]  
        \centering   
        \includegraphics[width=0.9\textwidth]{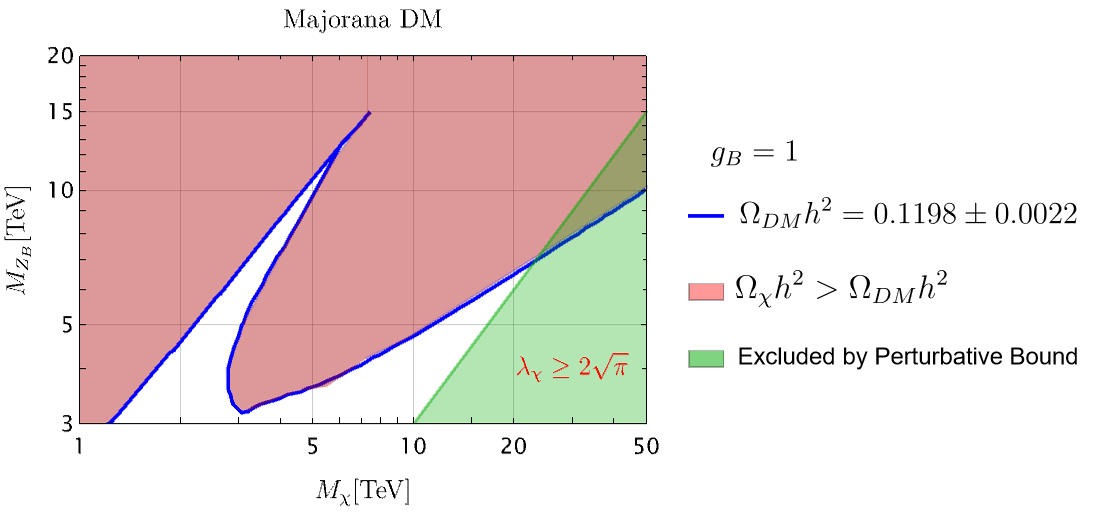}
         \caption{Allowed parameter space in the $M_{Z_B}-M_\chi$ plane when $g_B=1$. Here the blue solid line represents $\Omega h^2 = 0.1198\pm 0.0022$ and the red shaded region represents the parameter space when the relic density is greater than 0.12 and the green shaded region is excluded by the perturbative bound on $\lambda_\chi$. For illustration we use $M_{h_B}= 250$ GeV.}
         \label{DM-1.0}
\end{figure}    
\bibliography{refs}

\end{document}